\documentclass[useAMS,usenatbib]{mn2e}
\usepackage[usenames,dvipsnames]{color}
\usepackage{graphicx}
\usepackage{natbib}
\usepackage{subfigure}
\usepackage{mn2e-breakabs}
\usepackage{amsmath}
\usepackage{amssymb}    
\usepackage{lineno}

\long\def\symbolfootnote[#1]#2{\begingroup%
\def\thefootnote{\fnsymbol{footnote}}\footnote[#1]{#2}\endgroup} 

\def\TOTALCLUSTERNUMBERSpecHOD{174}       
\def\DRinBOSS{121}                        
\def\DRinBOSSzMin{0.203}                  
\def\DRinBOSSzMax{0.686}                  
\def\DRinBOSSTxMin{0.35}                  
\def\DRinBOSSTxMax{9.41}                    
\def\ANCinBOSS{53}                        
\def\ANCinBOSSzMin{0.207}                 
\def\ANCinBOSSzMax{0.699}                 
\def\ANCinBOSSTxMin{1.13}                 
\def\ANCinBOSSTxMax{10.37}                
\def\TOTALCLUSTERNUMBER{174}  
\def\XCSinCMASSft{74}                   
\def\DRSpecCMASSft{56}                   
\def\ANCSpecCMASSft{18}                   
\def\XCSinLOWZft{100}                    
\def\DRSpecLOWZft{65}                    
\def\ANCSpecLOWZft{35}                   
\def\notBOSSSpecUpdatedPercent{2}    
\def\MinHODnumber{0}                     
\def\MaxHODnumber{21}                    
\def\DRSpecCMASSfthonzeroText{three}     
\def\DRSpecCMASSfthonzeroTextCaps{Three}     
\def\MassBinsNumberText{five}            
\def\XCSinCMASSPhotft{76}                   
\def\NumExtraClustersOldCMASS{12}

\def\DRCMASSPhotft{61}            
\def\ANCCMASSPhotft{15}       
\def\TotalCMASSPhotft{76}     
\def\MinPhotHODnumber{0}            
\def\MaxPhotHODnumber{10.7}

\title[The HOD of BOSS galaxies in X-ray clusters]{The \textit{XMM} Cluster Survey: The Halo Occupation Number of BOSS galaxies in X-ray clusters}

\author[N. Mehrtens et al.]
{Nicola Mehrtens$^{1,2}$\thanks{E-mail: n.mehrtens@gmail.com}, 
A. Kathy Romer$^2$, 
Robert C. Nichol$^3$, 
Chris A. Collins$^4$, 
\newauthor  Martin Sahl\'en$^{5}$,
Philip J. Rooney$^2$,
Julian A. Mayers$^2$, 
A. Bermeo-Hernandez$^2$,
\newauthor Martyn Bristow$^4$,
Diego Capozzi$^3$,
L. Christodoulou$^{3}$,
Johan Comparat$^{6}$,
\newauthor 
Matt Hilton$^{7}$, 
Ben Hoyle$^{8}$,
Scott T. Kay$^{9}$,
Andrew R. Liddle$^{10}$,
Robert G. Mann$^{10}$,
\newauthor
Karen Masters$^{3}$,
Christopher J. Miller$^{11}$,
John K. Parejko$^{12}$,
Francisco Prada$^{6, 13, 14}$,
\newauthor
Ashley J. Ross$^{3,15}$,
Donald P. Schneider$^{16,17}$,
John P. Stott$^{5}$,
Alina Streblyanska$^{18}$,
\newauthor
Pedro T. P. Viana$^{19,20}$, 
Martin White$^{21,22}$, 
Harry Wilcox$^3$,
Idit Zehavi$^{23}$\\
$^{1}$~George P. and Cynthia W. Mitchell Institute for Fundamental Physics and Astronomy, Department of Physics and Astronomy,\\
Texas A\&M University, College Station, TX 77843, USA\\
$^{2}$~Astronomy Centre, University of Sussex, Falmer, Brighton, BN1 9QH, UK\\
$^{3}$~Institute of Cosmology and Gravitation, University of Portsmouth, Dennis Sciama Building, Portsmouth, PO1 3FX, UK \\
$^{4}$~Astrophysics Research Institute, Liverpool John Moores University, IC2, Liverpool Science Park, Brownlow Hill, Liverpool, L5 3AF \\
$^{5}$~BIPAC, Department of Physics, University of Oxford, Denys Wilkinson Building, 1 Keble Road, Oxford OX1 3RH, UK \\
$^{6}$~Instituto de F\'{\i}sica Te\'orica, (UAM/CSIC), Universidad Aut\'onoma de Madrid,  Cantoblanco, E-28049 Madrid, Spain \\
$^{7}$~Astrophysics \& Cosmology Research Unit, School of Mathematics, Statistics \& Computer Science, University of KwaZulu-Natal, \\ Westville Campus, Durban 4041, South Africa \\
$^{8}$~Universitaets-Sternwarte, Fakultaet fuer Physik, Ludwig-Maximilians Universitaet Muenchen, \\
Scheinerstr. 1, D-81679 Muenchen, Germany \\
$^{9}$~Jodrell Bank Centre for Astrophysics, School of Physics and Astronomy, The University of Manchester, Manchester M13 9PL, UK\\
$^{10}$~Institute for Astronomy, University of Edinburgh, Royal Observatory, Blackford Hill, Edinburgh, EH9 3HJ, UK\\
$^{11}$~Astronomy Department, University of Michigan, Ann Arbor, MI 48109, USA \\
$^{12}$~Department of Astronomy, University of Washington, Box 351580, Seattle, WA 98195, USA\\
$^{13}$~Instituto de Astrof\'{\i}õsica de Andaluc\'{\i}a (CSIC), Glorieta de la Astronom\'{\i}a, E-18080 Granada, Spain\\
$^{14}$~Instituto de Fisica Teorica, Universidad Autonoma de Madrid,  Cantoblanco E-28049, Madrid, Spain\\
$^{15}$~Center for Cosmology and AstroParticle Physics, The Ohio State University, Columbus, OH 43210, USA\\
$^{16}$~Department of Astronomy and Astrophysics, The Pennsylvania State University, University Park, PA 16802\\
$^{17}$~Institute for Gravitation and the Cosmos, The Pennsylvania State University, University Park, PA 16802\\
$^{18}$~Instituto de Astrof\'{\i}sica de Canarias (IAC), Calle V\'{\i}a Lactea, s/n, 38200, La Laguna, Tenerife, Spain\\
$^{19}$~Instituto de Astrof\'{\i}sica e Ci\^{e}ncias do Espa\c{c}o, Universidade do Porto, CAUP, Rua das Estrelas, 4150-762 Porto, Portugal \\
$^{20}$~Departamento de F\'isica e Astronomia, Faculdade de Ci\^encias, Universidade do Porto, Rua do Campo Alegre, 687, \\ 4169-007 Porto, Portugal \\
$^{21}$~Departments of Physics and Astronomy, University of California, Berkeley, CA 94720, USA \\ 
$^{22}$~Lawrence Berkeley National Laboratory, 1 Cyclotron Road, Berkeley, CA 94720, USA \\
$^{23}$~Department of Astronomy, Case Western Reserve University, 10900 Euclid Avenue, Cleveland, OH 44106, USA}

\date{Accepted for publication in MNRAS}
\pubyear{2016}

\begin{document}
\label{firstpage}
\pagerange{\pageref{firstpage}--\pageref{lastpage}}
\maketitle

\begin{abstract}

We present a direct measurement of the mean halo occupation distribution (HOD) of galaxies taken from the eleventh data release (DR11) of the Sloan Digital Sky Survey-III Baryon Oscillation Spectroscopic Survey (BOSS). The HOD of BOSS low-redshift (LOWZ: $0.2 < z < 0.4$) and Constant-Mass (CMASS: $0.43 <z <0.7$) galaxies is inferred via their association with the dark-matter halos of \TOTALCLUSTERNUMBERSpecHOD\, X-ray-selected galaxy clusters drawn from the \textit{XMM} Cluster Survey (XCS). Halo masses are determined for each galaxy cluster based on X-ray temperature measurements, and range between ${\rm log_{10}} (M_{180}/M_{\odot}) = 13-15$. Our directly measured HODs are consistent with the HOD-model fits inferred via the galaxy-clustering analyses of Parejko et al. for the BOSS LOWZ sample and White et al. for the BOSS CMASS sample. Under the simplifying assumption that the other parameters that describe the HOD hold the values measured by these authors, we have determined a best-fit alpha-index of 0.91$\pm$0.08 and $1.27^{+0.03}_{-0.04}$ for the CMASS and LOWZ HOD, respectively. These alpha-index values are consistent with those measured by White et al. and Parejko et al. In summary, our study provides independent support for the HOD models assumed during the development of the BOSS mock-galaxy catalogues that have subsequently been used to derive BOSS cosmological constraints.

\end{abstract}


\section{Introduction}
\label{introduction}

In the hierarchical formation scenario, large-scale structures in the Universe arise through the successive mergers of increasingly-large dark matter halos. These haloes cannot be observed directly, but their presence can be inferred from the galaxies they contain, assuming light traces mass. Galaxy surveys therefore can be applied to studies of both  cosmology
(e.g., \citealt{2005MNRAS.362..505C,2005ApJ...633..560E,2007MNRAS.381.1053P,Blake:2011b,2012MNRAS.425..415S,2012ApJ...745...16T,2012PhRvD..86j3518P, 2014MNRAS.441...24A})
and galaxy evolution 
(e.g., \citealt{Tinker:2010,2010MNRAS.406.1306A,Zehavi:2011,2012ApJ...744..159L,2012MNRAS.424..232W,2012MNRAS.424.2339T}).

An essential component of many galaxy survey based cosmology and galaxy evolution studies is the Halo Occupation Distribution (HOD) model (e.g., \citealt{Peacock:2000,Seljak:2000,BerlindWeinberg:2002,CooraySheth:2002,2004ApJ...609...35K}). This model encapsulates the complicated physics of galaxy formation and evolution within a relatively simple framework. HOD describes the mean number of galaxies above a luminosity threshold within a virialised halo of given mass. Under the HOD framework, the number of galaxies populating a halo increases, on average, as a function of halo mass.  Galaxies populating a halo are divided into either `central' or `satellite' galaxies 
(e.g., \citealt{2003ApJ...593....1B,Zheng:2005,2005ApJ...633..791Z}). Depending on its mass and evolution history, a halo can host, or be devoid of, either or both types of galaxies (above the chosen luminosity threshold). 

Dark matter halos can accrete satellite galaxies and grow in mass through halo-halo mergers. The central (and satellite) galaxies of the newly acquired sub-halos become the satellite galaxies of the dominant halo. In HOD nomenclature, the `two-halo' term refers to the region of the HOD where the physical separation between galaxies is sufficiently large that the clustering statistic counts pairs of galaxies hosted by separate dark matter halos; whereas the `one-halo' term refers to the non-linear regime where the clustering statistic counts pairs of galaxies hosted by the same dark-matter halo.

Several methods have been implemented to measure the form of the HOD for a given galaxy type. These include fitting a model to the HOD predicted by galaxy-clustering analyses 
(e.g., \citealt{2005ApJ...625..613A,Zheng:2007,Zheng:2009,White:2011,Parejko:2013,Reid:2014,Nuza:2014,Guo:2015,2015ApJ...806....1M,2015ApJ...806....2M,
2016MNRAS.460.1173R}), measurements of the galaxy conditional luminosity function 
(e.g., \citealt{Yang:2003,Cooray:2006,VanDenBosch:2007,Yang:2008,
2013ApJ...767...92R,
2016MNRAS.459.3040G}), satellite kinematics, 
(e.g., \citealt{More:2009,More:2011}) galaxy-galaxy lensing, 
(e.g., \citealt{
2012ApJ...744..159L,
2015MNRAS.454.1161Z,
2015arXiv150705353P}) or by directly counting the number of galaxies within pre-determined dark-matter halos e.g., such as those identified by galaxy cluster/group surveys 
(e.g., \citealt{Lin:2004,CollisterLahav:2005,Ho:2009,2009ApJ...698..143R,Capozzi:2012,Tinker:2012,Hoshino:2015,
2012MNRAS.419.2821C}). 
  
The Sloan Digital Sky Survey-III Baryon Oscillation Spectroscopic Survey (or BOSS, \citealt{Eisenstein:2011}) is a spectroscopic survey that has measured redshifts for $\simeq$1.5 million galaxies over an area of  $\simeq$10,000\,deg$^{2}$. The primary scientific goal of BOSS is to place constraints on cosmological models by measuring the Baryon Acoustic Oscillation (BAO) feature \citep{PeeblesYu:1970,SunyaevZeldovich:1970,Doroshkevich:1978}. BOSS also enables other science, for example studies of galaxy evolution and galaxy bias. Using the galaxy-clustering approach, measurements of the HOD of BOSS galaxies have been presented in both \cite{White:2011} and \citet{Parejko:2013}. Using the first year of BOSS spectroscopic data, \citet{White:2011} performed a measurement of the real- and redshift-space clustering of BOSS CMASS-galaxies at $z\sim0.5$, and simultaneously fit an HOD model to these data to predict the mean number of CMASS-galaxies contained within a halo of given mass. A similar analysis, using low-redshift BOSS galaxies, was performed by \citet{Parejko:2013}, in which they predict the HOD of BOSS LOWZ-galaxies at $z\sim0.3$.

In this paper, we test the HOD models of \cite{White:2011} and \citet{Parejko:2013} (W11 and P13, respectively hereafter) by directly counting the number of BOSS galaxies in the vicinity of X-ray clusters taken from the {\textit XMM} Cluster Survey (XCS; \citealt{Romer:2001}) in the SDSS DR11 BOSS spectroscopic footprint \citep{Alam:2015}. Our motivation for this project is that the W11 and P13 HOD models have been adopted by many of the subsequent BOSS science analyses, and it is important to check them using an independent technique. 

Clusters selected using optical/near-IR galaxy over density methods suffer from mis-centering issues, e.g. \citealt{2016ApJS..224....1R}, that could impact HOD measurements. Therefore, we have chosen X-ray selected clusters for this study. In principle, we would like to have weak lensing mass measurements for all the clusters in our sample. However, in practice, it is not yet possible gather the required data for large numbers of clusters: the largest recent studies are limited to $\simeq$ 50 clusters, e.g. \citealt{2016MNRAS.456L..74S,2016MNRAS.457.1522A}. Therefore, for this study, we have used cluster-averaged X-ray temperatures, ${\it T}_{\rm X}$, combined with an externally calibrated $T_{\rm X}$--$M$ relation. (The scatter on the $T_{\rm X}$--$M$ is predicted, using simulations, to be $< \simeq 20\%$, \citealt{2006ApJ...650..128K}). Previous X-ray based HOD studies (e.g. \citealt{Lin:2004,Ho:2009}) relied on X-ray luminosity (${\it L}_{\rm X}$) as their mass proxy because they did not have access to large numbers of homogeneously derived ${\it T}_{\rm X}$ values. We also note that the spatial resolution of {\it XMM-Newton} precludes the measured of core excised ${\it L}_{\rm X}$ values at the redshift of most of our clusters. (It is only after core excision that ${\it L}_{\rm X}$ can be used as a reliable mass proxy, e.g., \citealt{Stanek:2006,Maughan:2007}). 

The structure of the paper is as follows: Section~\ref{The data}, describes the BOSS galaxy redshift catalogues, and X-ray cluster samples used in the analysis, as well as the methods used to estimate virial masses, virial radii, velocity dispersions, redshifts and X-ray temperatures for the clusters. Section~\ref{Method} presents the HOD measurements. Section~\ref{HODCompIntro}, compares those measurements to the W11 and P13 HOD models. Section~\ref{Discussion} discusses the implications of our findings, and possible sources of systematic error. Throughout, we assume a flat $\Lambda$CDM cosmology with values $\Omega_{\rm m}=0.274$, $\Omega_{\Lambda}=0.726$ and $h=0.7$ (as used in W11 and P13). Co-moving separations are measured in $h^{-1}$\,Mpc, with $H_{0}=100h\,\rm{km~s}^{-1}$\,Mpc$^{-1}$. In the following, when we refer to dark matter halos in the `one-halo' regime, we mean those of sufficient mass that they could contain satellite galaxies. For the redshifts considered in our analysis,  the `one-halo' regime typically applies to halos of mass ${\rm log}_{10} M_{180} \sim 13-15 M_{\odot}$ (where $M_{180}$ is the mass contained in a spherical overdensity $\Delta_{180}$ with radius $R_{180}$).

\section{The data}
\label{The data}

The data used in this paper is taken from two main sources; the Baryon Oscillation Spectroscopic Survey (BOSS; \citealt{Dawson:2013}, Section~\ref{BOSSdata}) and the \textit{XMM} Cluster Survey (XCS; \citealt{Romer:2001}, Section~\ref{XCSdata}).  

\subsection{BOSS Data}
\label{BOSSdata}

The third phase of the Sloan Digital Sky Survey (SDSS; \citealt{York:2000}), termed SDSS-III (\citealt{Eisenstein:2011}), included four projects: BOSS \citep{Eisenstein:2011}, SEGUE-2 \citep{SEGUE:2010}, MARVELS \citep{MARVELS:2011}, APOGEE \citep{APOGEE:2013}.  Data was obtained using the 2.5-m Sloan telescope \citep{Gunn:2006} at Apache Point Observatory, New Mexico, and the SDSS spectrographs \citep{Smee:2013}. The study presented here only makes use of BOSS data products \citep{Bolton:2012,Dawson:2013}.

BOSS was designed to measure the BAO feature at $z \leq 0.7$ to sub 2 per cent accuracy using luminous galaxies with an approximately constant co-moving number density ($\bar{n}\simeq3 \times 10^{-4} h^{3}$~$ \rm Mpc^{-3}$). Galaxies were selected for BOSS spectroscopic observation from $u$, $g$, $r$, $i$, $z$ imaging data \citep{Fukugita:1996,Gunn:1998} taken from the Eight and Ninth SDSS Data Releases (SDSS DR8, \citealt{Aihara:2011}; SDSS DR9, \citealt{Eisenstein:2011}). The galaxy targets were selected using colour and magnitude cuts that track the expected evolution of passively evolving luminous red galaxies (LRGs) with redshift. These evolutionary tracks are based on the population synthesis models of \citet{Maraston:2009}. Due to the transition of the 4000\,\AA\, break of LRGs between the $g$ and $r$ filters at $z\sim0.4$, two sets of cuts were necessary. This divided the targets into two broad redshift bins; a low-redshift sample (LOWZ; Equation~\ref{Eqn:LOWZ}) spanning the redshift range $0.2 \leq z \leq 0.4$, and a high-redshift sample (CMASS - for ``Constant Mass''; Equation~\ref{Eqn:CMASS}) spanning the redshift range $0.43 \leq z \leq 0.7$. The CMASS sample is defined by:

\begin{equation}
r < 13.6 + c_{\parallel}/0.3, |c_{\perp}| < 0.2, 16 < r < 19.5, \label{Eqn:LOWZ}
\end{equation}

\begin{equation}
d_{\perp} > 0.55, i < 19.86 + 1.6 \times (d_{\perp} - 0.8), 17.5 < i < 19.9,\label{Eqn:CMASS}
\end{equation}

\noindent where the colours $c_{\parallel}$, $c_{\perp}$ and $d_{\parallel}$ are given by Equations~\ref{Eqn:ColCut1},\ref{Eqn:ColCut2}, and \ref{Eqn:ColCut3}, respectively.

\begin{equation}
c_{\parallel} = 0.7 \times (g - r) + 1.2 \times (r - i - 0.18), \label{Eqn:ColCut1}
\end{equation}

\begin{equation}
 c_{\perp} = (r - i) - (g - r)/4 - 0.18, \label{Eqn:ColCut2}
\end{equation}

\begin{equation}
d_{\perp} = (r - i) - (g - r)/8. \label{Eqn:ColCut3}
\end{equation}

Compared to the SDSS-I/II spectroscopic survey of LRGs \citep{Eisenstein:2001}, the BOSS colour cuts extend to intrinsically bluer colours and fainter magnitudes.  As a result, BOSS targets consist of luminous galaxies, rather than luminous {\it red} galaxies. The emphasis is on constant stellar mass rather than on a particular galaxy type; for example, Equation~\ref{Eqn:ColCut2} results in a sample that is effectively volume limited to $z \sim 0.6$, and approximately stellar mass limited to $z=0.7$. These properties have been confirmed by  \citet{Masters:2011} who studied HST\footnote{{\tt http://hubblesite.org}} images of 240 BOSS targets that lay in the  COSMOS survey area. They demonstrate that 23 per cent of the BOSS targets are late-type, star-forming, galaxies. Only by employing an additional, $g-i >2.35$, colour cut were \citet{Masters:2011} able to produce a sub-sample reminiscent of the SDSS I/II-LRG sample, i.e., one containing more than 90 per cent early-type galaxies. Not limiting BOSS targets to a particular colour/morphological type provides a more representative census of the galaxy population within the desired stellar mass range. This feature is important because it has been shown that galaxies of different type cluster differently 
\citep{Simon:2009, 2009MNRAS.398..807S, Zehavi:2011, 2014ApJ...784..128S} 
and BOSS aims to probe luminosity--dependent clustering.

In this study, we draw on the spectroscopic data that were released (SDSS DR11) internally to BOSS collaborators on July 3rd, 2013, i.e., before the public data release in January 2015 (DR12). DR11 and DR12 are described in \citet{Alam:2015}. DR11 includes spectra obtained through to July 2013 and covers 7,341\,deg$^{2}$ of the sky. By comparison, DR12 contains additional spectra and covers more area (10,400\,deg$^{2}$). Not all of the redshifts included in DR11 were gathered during SDSS-III: some were obtained during earlier (i.e. SDSS-I/II) campaigns, and some were assigned using the {\it close-pair correction} technique. With regard to the latter, for the SDSS-III spectrographs \citep{Smee:2013}, fibre collision occurs at angular separations smaller than 62\,arcsecs. This is a problem for BOSS because luminous galaxies tend to be strongly clustered on the sky. Therefore, in the event that two galaxies -- that have been selected as potential targets for SDSS-III spectroscopy -- are within 62\,arcsecs of one another, the one that is not allocated a fibre is assigned the spectroscopic redshift of its neighbour (see \citealt{Dawson:2013} for more information). 

Because DR11 did not include a complete spectroscopic census of all of the galaxies selected to be BOSS targets, we use the terms ``BOSS-target'', ``BOSS-galaxy'', ``CMASS-galaxy'',  and ``LOWZ-galaxy'', in this paper. The BOSS-target sample is the superset, the BOSS-galaxy sample is the subset of these with redshift information (spectroscopic or close-pair corrected) in DR11. The BOSS-galaxy sample is the union of the distinct CMASS-galaxy and LOWZ-galaxy samples.

\subsection{XCS data}
\label{XCSdata}

XCS uses all available data in the {\it XMM-Newton} ({\it XMM}) public archive to search for galaxy clusters that were detected serendipitously in {\it XMM} images. 
X-ray sources are detected in {\it XMM} images using an algorithm based on wavelet transforms (see \citealt{Lloyd-Davies:2011}, henceforth LD11, for details). 
Sources are then compared to a model of the instrument point spread function to determine if they are extended: XCS uses the signature of X-ray extent to distinguish clusters from more common X-ray sources, such as active galactic nuclei.  Optical imaging is used to confirm the identity of the extended sources (most are clusters, but low-redshift galaxies and supernova remnants are also extended in {\it XMM} images). Where possible, either a photometric or spectroscopic redshift is determined for the confirmed cluster. For each confirmed cluster with an associated redshift, cluster-averaged X-ray luminosities (${\it L}_{\rm X}$), and cluster-averaged X-ray temperatures (${\it T}_{\rm X}$) are measured using an automated pipeline (LD11). 

The majority of the X-ray clusters used in the HOD study described herein were drawn from the first XCS data release (XCS-DR1; \citealt{Mehrtens:2012}, henceforth M12), with the remainder from the ``XCS-Ancillary'' sample, see below. The optical imaging and spectroscopic campaign described in M12 resulted in 503 optically-confirmed clusters, including 464 with redshifts, and 401 with ${\it T}_{\rm X}$ estimates. 
The XCS-DR1 clusters are distributed across the sky and span the redshift and temperature ranges $0.06 < z < 1.46$ and $0.04\, \rm \,keV  < {\it T}_{\rm X} <  14.7\, \rm \,keV$, respectively. 
This temperature range corresponds to halo masses  of ${\rm log_{10}} (M_{180}/M_{\odot}) = 13-15$ (see Section~\ref{ClusterMasses}). 
For our HOD study, we used \DRinBOSS\, XCS-DR1 clusters that are located within the spectroscopic BOSS footprint. The footprint has a complex shape, so the \textsc{mangle} software \citep{Swanson:2008} was used to track its angular completeness, using a completeness threshold of 0.8 for a cluster to be included in the spectroscopic footprint. These \DRinBOSS\, XCS-DR1 clusters have redshifts and temperatures in the range $ \DRinBOSSzMin\, <z< \DRinBOSSzMax\,$ and  $\rm \DRinBOSSTxMin\, < {\it T}_{\rm X} <  \DRinBOSSTxMax\, \rm \,keV$.

The XCS-Ancillary sample includes extended {\it XMM} sources that were not included in XCS-DR1 (M12) for one of three reasons: they were associated with the target of the respective {\it XMM} image (and hence were not serendipitous detections); they were not included in any of the three XCS-Zoo exercises (used to optically confirm the XCS-DR1 clusters, see Section~4 of M12); or they were detected in {\it XMM} images processed in the time elapsed since the publication of M12. Initial redshifts were assigned to these clusters using NED\footnote{http://ned.ipac.caltech.edu} identifications using the method described in Section~4.1 of LD11. These redshifts were refined using the method described in Section~\ref{betterTzs}.  For our HOD study, we used \ANCinBOSS\, XCS-Ancillary clusters that are located within the spectroscopic BOSS footprint. These  \ANCinBOSS\, clusters have redshifts and temperatures in the range $\ANCinBOSSzMin\, <z<\ANCinBOSSzMax\,$ and  $\ANCinBOSSTxMin\,< {\it T}_{\rm X} <\ANCinBOSSTxMax\, \rm \,keV$.

\subsection{Cluster velocity dispersions, masses, and radii}
\label{ClusterMasses}

In order to validate cluster redshifts (see Section~\ref{betterTzs}), and to measure halo occupation numbers (see Section~\ref{Method}), we need to estimate cluster velocity dispersions, masses and radii. For the velocity dispersions, we use the empirical $\sigma_{v}- T_{\rm X}$ relation of \citet{XueWu:2000}. This relation is based on a sample of 145 X-ray groups and clusters with temperatures ranging from $ 0.1\,{\rm keV} < T_{\rm X} < 10\,{\rm keV} $, and is therefore similar to the X-ray temperature range of the clusters in XCS-DR1. Several fits are presented in \citet{XueWu:2000}; we have chosen to adopt the relation measured via an orthogonal distance regression fit to the whole sample\footnote{\citet{XueWu:2000} note that the cluster and group relations when fitted separately are consistent with the combined fit.}, because this method accounts for uncertainties in both the $\sigma_{v}$ and $T_{\rm X}$ values. The relation is given by
\begin{equation}
 \sigma_{v} = 10^{2.51 \pm 0.01} T_{\rm X}^{0.61 \pm 0.01},
\end{equation} 
where $\sigma_{v}$ and $T_{\rm X}$ are in units of $\rm km s^{-1}$ and keV, respectively.

For the cluster masses, we adopt the prescription in \citet{Sahlen:2009}, which involves fitting the following model to each cluster:
\begin{equation}
T_{\rm X} = T_{\rm X, mean}(M_{180}) + \Delta_{\log T_{\rm X}}\,,
\end{equation}
where $T_{\rm X, mean}(M_{180})$ is the mean $M$--$T_{\rm X}$ relation and $\Delta_{\log T_{\rm X}}$ represents the scatter of individual clusters about the mean relation. 

We parameterize the $M$--$T_{\rm X}$ relation according to the self-similar prediction \citep[e.g.,][]{Kaiser:1986,Bryan:1998,Voit:2005},
\begin{equation}
T_{\rm X, mean} = A M_{\rm vir}^{2/3}\,[\Delta_{\rm vir}(z)E^{2}(z)]^{1/3}\,,
\end{equation}
where $M_{\rm vir}$ is the virial mass of the cluster, $\Delta_{\rm vir}(z)$ the spherical overdensity within the virial radius of the cluster, and $E^{2}(z)$ is the reduced Hubble parameter for our cosmological model,
\begin{equation}
  E^{2}(z) = \Omega_{\rm m}(1+z)^{3} + \Omega_{\rm k}(1+z)^{2} + \Omega_{\Lambda}\,,
\end{equation}
 with $\Omega_{\rm k} = 1 - \Omega_{\rm m} - \Omega_{\Lambda}$. In our analysis, we restrict ourselves to a flat universe with $\Omega_{\rm k}=0$. $A$ is a normalization constant set by requiring 
\begin{equation}
M_{500} = 3\times10^{14}\,h^{-1}\,{\rm M_{\sun}}\,,
\end{equation}
at $z=0.05$ for $T_{\rm X}=5$\,keV. Our fiducial cosmological model reproduces the local abundance of galaxy clusters as given by the HIFLUGCS catalogue \citep{Pierpaoli:2001, Reiprich:2002, Viana:2003}.

Conversions between $M_{180}$, $M_{500}$ and $M_{\rm vir}$ are performed using the standard \citet{NavarroFrenkWhite:1996} profile prescription by \cite{Hu:2003} with a halo concentration parameter of $c=5$. 
We have tested the impact of changing the concentration parameter on the mass and radius estimates, and find that the change, compared to $c=5$, to the mean value is much less than the one sigma errors, when using either $c=2.5$ or $c=10$.

We model the scatter $\Delta_{\log T_{\rm X}}$ as log-normal about the mean $M$--$T_{\rm X}$ relation, with a standard deviation $\sigma_{\log T_{\rm X}} = 0.1$. This model is motivated by observational estimates of the instrinsic scatter (e.g., \citealt{Arnaud:2005,Kravtsov:2006,Zhang:2006}) and results from N-body hydrodynamic simulations (e.g., \citealt{Viana:2003,Borgani:2004,Balogh:2006,Kravtsov:2006}). The likelihood is constructed from the $T_{\rm X}$ measurement probability distributions, modelled by a split normal distribution: 
\begin{equation}
\mathcal{L}(T_{\rm X}) = A \exp(-(T_{\rm X}-T^{*}_{\rm X})^2/2\sigma^2)\,,
\end{equation}
where

\begin{equation}
\sigma = 
\begin{cases}
\sigma_+ & \mbox{if } T_{\rm X} \geq T^{*}_{\rm X} \\
\sigma_- & \mbox{if } T_{\rm X} < T^{*}_{\rm X},
\end{cases}
\end{equation}
with $A = \sqrt{2/\pi}(\sigma_+ + \sigma_-)^{-1}$. Here, $T^{*}_{\rm X}$ is the measured central value of $T_{\rm X}$, $\sigma_+$ and $\sigma_-$ the upper and lower 1-$\sigma$ uncertainties. The likelihood is explored using Monte Carlo Markov Chain (MCMC, e.g. \citealt{Gelman:1992, LewisBridle:2002, Tegmark:2004, Dunkley:2005}) sampling, using $M_{180}$ and the temperature scatter $\Delta_{\log T_{\rm X}}$ as free parameters.  An uninformative flat prior is placed on $M_{180}$, and a prior $N(0,\sigma_{\log T_{\rm X}})$ on $\Delta_{\log T_{\rm X}}$.  Approximately 10,000 Markov chain elements are generated for each cluster, for which the distributions have converged. In addition to visual inspection of chain statistics, we assess this using the Gelman-Rubin test. We require the Gelman-Rubin ratio $R < 1.05$.

The cluster radii ($R_{180}$) are derived parameters in the MCMC procedure, computed by assuming that the cluster mass $M_{180}$ is contained in a spherical overdensity $\Delta_{180}$ with radius $R_{180}$. The MCMC procedure thereby produces chain samples of the distributions of $R_{180}$ values. From these samples, the mean $R_{180}$ values and their uncertainties are derived. 

\subsection{Cluster redshifts and temperatures}
\label{betterTzs}

\begin{figure}
\begin{center}
{    
\includegraphics[scale=0.4]{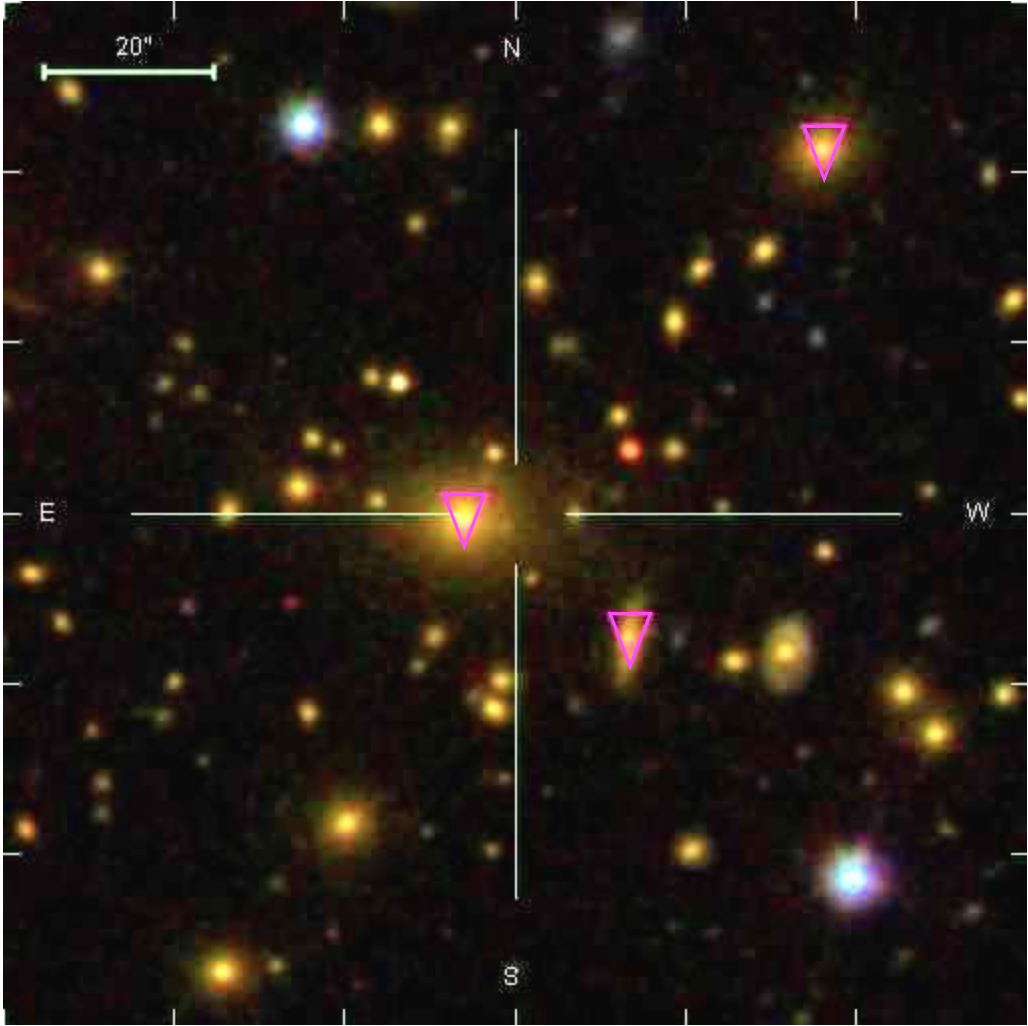}
}
\begin{subfigure}
{
\includegraphics[scale=0.4]{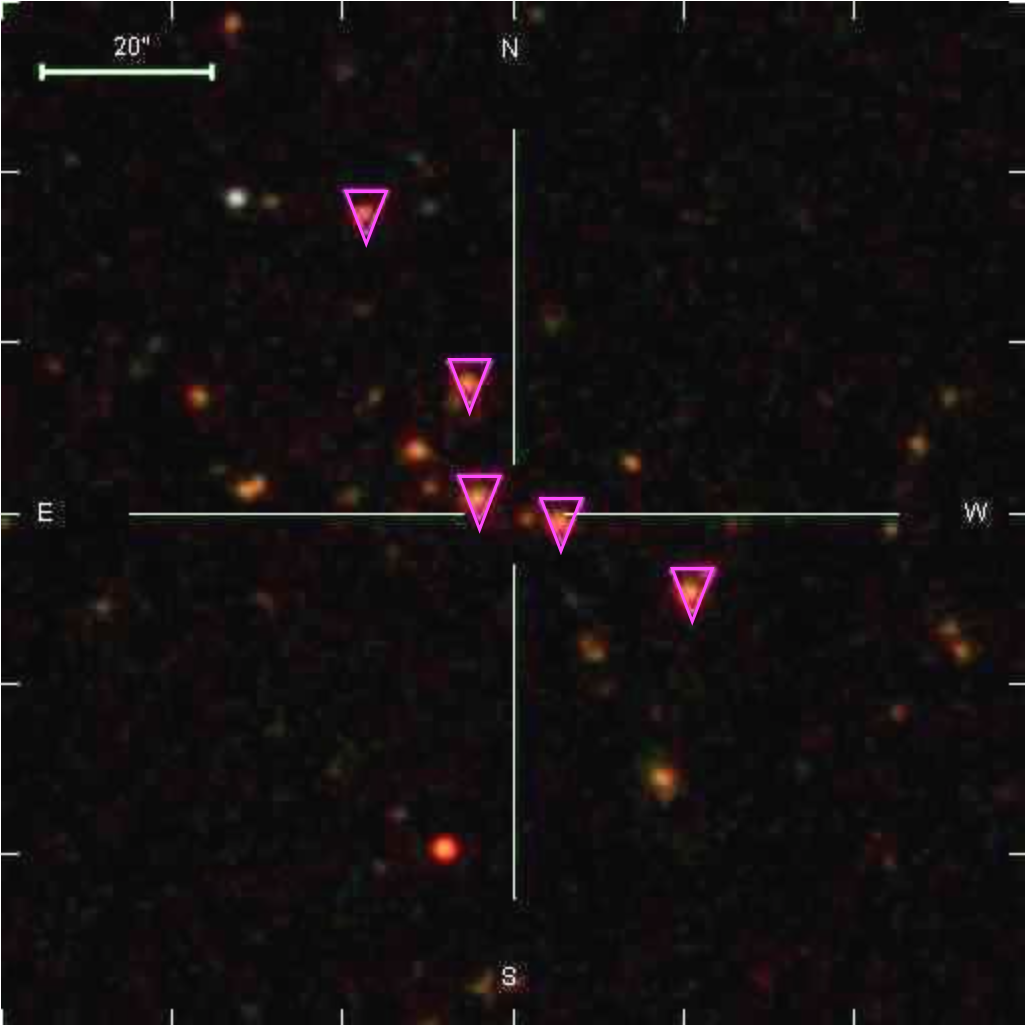}
}
\end{subfigure}
\\
\caption{The clusters XMMXCS J133453.1$+$405654.5 (top; $z=0.233$) and XMMXCS J112259.3$+$465916.8 (bottom: $z=0.480$) as imaged by SDSS DR8. False colour-composite images show the central $2 \times 2$ arcmin region of each cluster. Highlighted by pink triangles are SDSS DR11 BOSS-galaxies falling within a projected $R_{180}$ radius. These were adopted as member galaxies and used to assign a spectroscopic cluster redshift based on BOSS spectroscopy.}  
\label{SDSSimageCompilation}
\end{center}
\end{figure}

For many clusters in our study we have updated the M12 (XCS-DR1) or NED (XCS-Ancilliary) redshift estimates using BOSS spectroscopy. For this, galaxy redshifts were extracted from the appropriate BOSS spectroscopic redshift catalogue.  See Fig.~\ref{SDSSimageCompilation} for examples.
 
For clusters with spectroscopic redshifts in M12 or NED, we defined a cylinder centred on the XCS centroid. This cylinder had a radius, on the sky, of $R_{180}$. The cylinder had a depth, along the line of sight, of $\pm 3\sigma_{v}$. (The $R_{180}$ and $\sigma_{v}$ values were estimated from the $T_{\rm X}$ values using the method described in Section~\ref{ClusterMasses}). Spectroscopic redshifts for the galaxies enclosed by the cylinder were then extracted from the BOSS catalogues. For clusters with only a photometric redshift in M12 or NED, we again defined a cylinder centred on the XCS centroid with a radius of $R_{180}$, but this time set no bounds along the line of sight. 

If more than one BOSS redshift was extracted for a given cylinder, we followed the redshift-gapper method (\citealt{Halliday:2004});  i.e., we identify the location of the most likely peak in redshift space and determine the mean cluster redshift using all galaxies with redshifts within $\Delta z=0.015$ of that peak.

If no galaxies were extracted for a given cylinder, then the respective cluster was still included the HOD analysis (Section~\ref{Method}) if it had a spectroscopic redshift in M12 or NED. After all, BOSS targets represent only a subset of galaxies, i.e., there will be other cluster members that do not meet the colour and magnitude thresholds described in Section~\ref{BOSSdata}. However, if the cluster only had a photometric redshift in M12 or NED, it was excluded from the HOD analysis. We discuss possible implications of this approach in Section~\ref{HODPhot}.

After the CMASS- or LOWZ-galaxies were extracted for the respective cluster, the SDSS image was inspected to check the location of the galaxies with respect to the X-ray emission (see Fig.~\ref{SDSSimageCompilation}). This highlighted the fact that the XCS-DR1 spectroscopic redshift of XMMXCS J023346.0$-$085048.5 was inaccurate: in M12 it had been  based on the observation of a single galaxy that yielded a redshift of $z=0.25$. The same galaxy was measured to have a redshift of $z=0.265$ by BOSS (the BOSS value was adopted for the cluster). In this case, defining a cylinder based on the M12 redshift did not automatically extract the BOSS redshift for that galaxy.

\begin{figure}
{    
\includegraphics[scale=0.55]{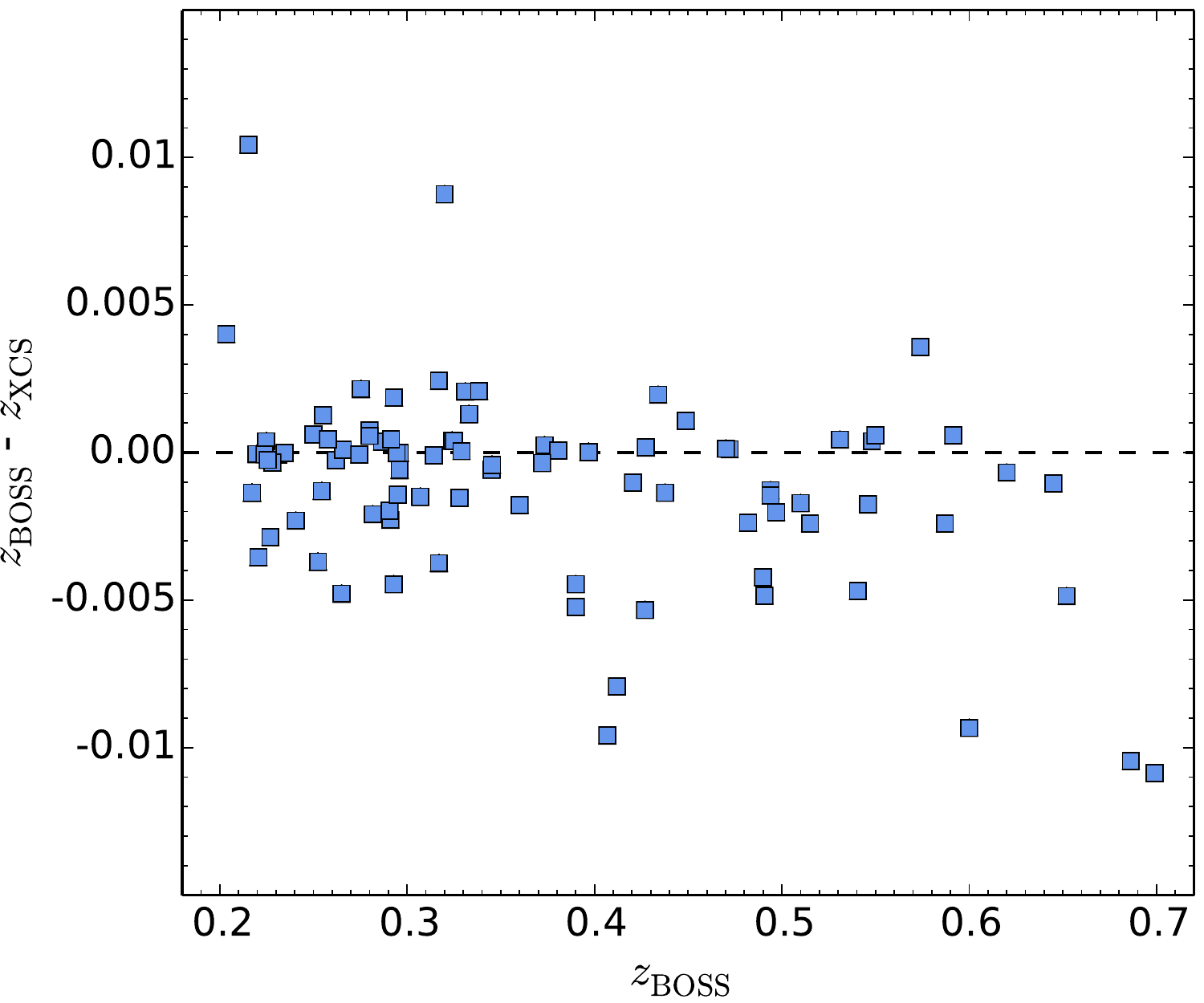}
}
\begin{subfigure}
{
\includegraphics[scale=0.55]{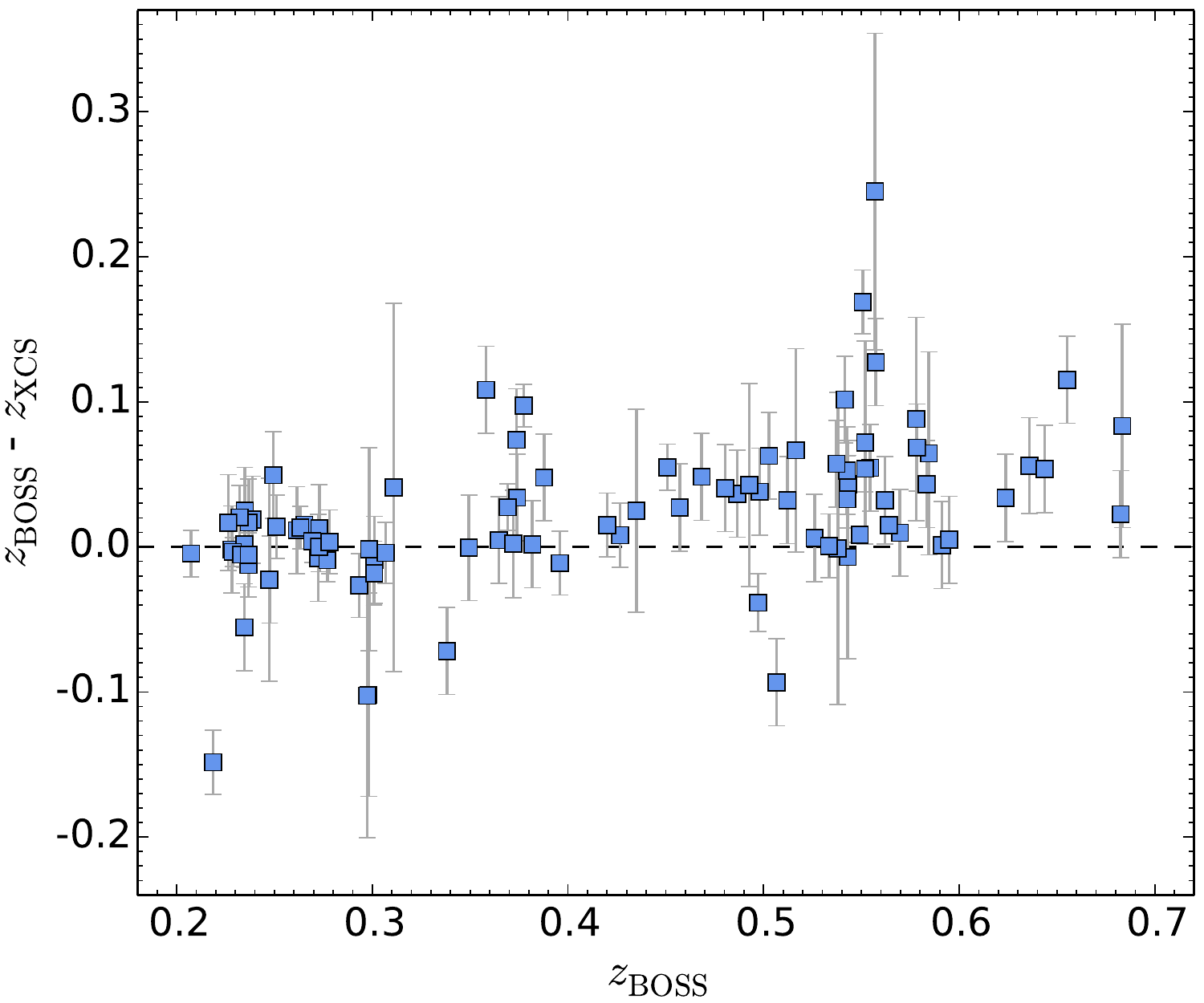}
}
\end{subfigure}
\\
\caption{Comparison between XCS-DR1 cluster redshifts published in M12, $z_{\rm XCS}$, and the updated values based on BOSS spectroscopy, $z_{\rm BOSS}$. TOP: Clusters with spectroscopic redshifts in M12. The 1-$\sigma$ dispersion is $\Delta_{z}=0.003$. BOTTOM: Clusters with photometric redshifts in M12. The 1-$\sigma$ dispersion is $\Delta_{z}=0.05$.}
\label{BOSSzDR1zComp}
\end{figure}

The resulting changes were small ($\Delta z < 0.02$) when spectroscopically-determined cluster redshifts were used as the input; see Fig.~\ref{BOSSzDR1zComp} (top). However, as expected, they were larger when photometric redshifts were used as inputs; the changes ranged up to $\Delta z=0.25$, although 90 per cent were less than $|\Delta z=0.1|$; see Fig.~\ref{BOSSzDR1zComp} (bottom). From Fig.~\ref{BOSSzDR1zComp} (bottom), it is clear that above $z_{\rm BOSS}\simeq0.3$, the photometric redshift estimates in M12 are systematically low. The same effect was highlighted in M12 (see discussion in Section~5.3 of M12).

There is a known degeneracy between $z$ and $T_{\rm X}$ in X-ray spectral fitting (e.g. see \citealt{Liddle:2001}), so we remeasured the $T_{\rm X}$ values once the BOSS determined clusters redshifts were in hand.  The method used is as described in LD11, but using updated {\textit XMM} calibration and XSPEC (12.8.1g) versions. Specifically,  we have fitted the {\textit XMM} spectra to a WABS$\times$MEKAL model \citep{MeweSchrijver:1986}, fixing the Hydrogen column density to the \citet{Dickey:1990} value and the metal abundance to 0.3 times the Solar value. For consistency,  $T_{\rm X}$ values were also re-measured using the updated {\textit XMM} calibration and XSPEC versions even if the redshifts had not changed compared to M12.

\begin{figure}
\includegraphics[scale=0.55]{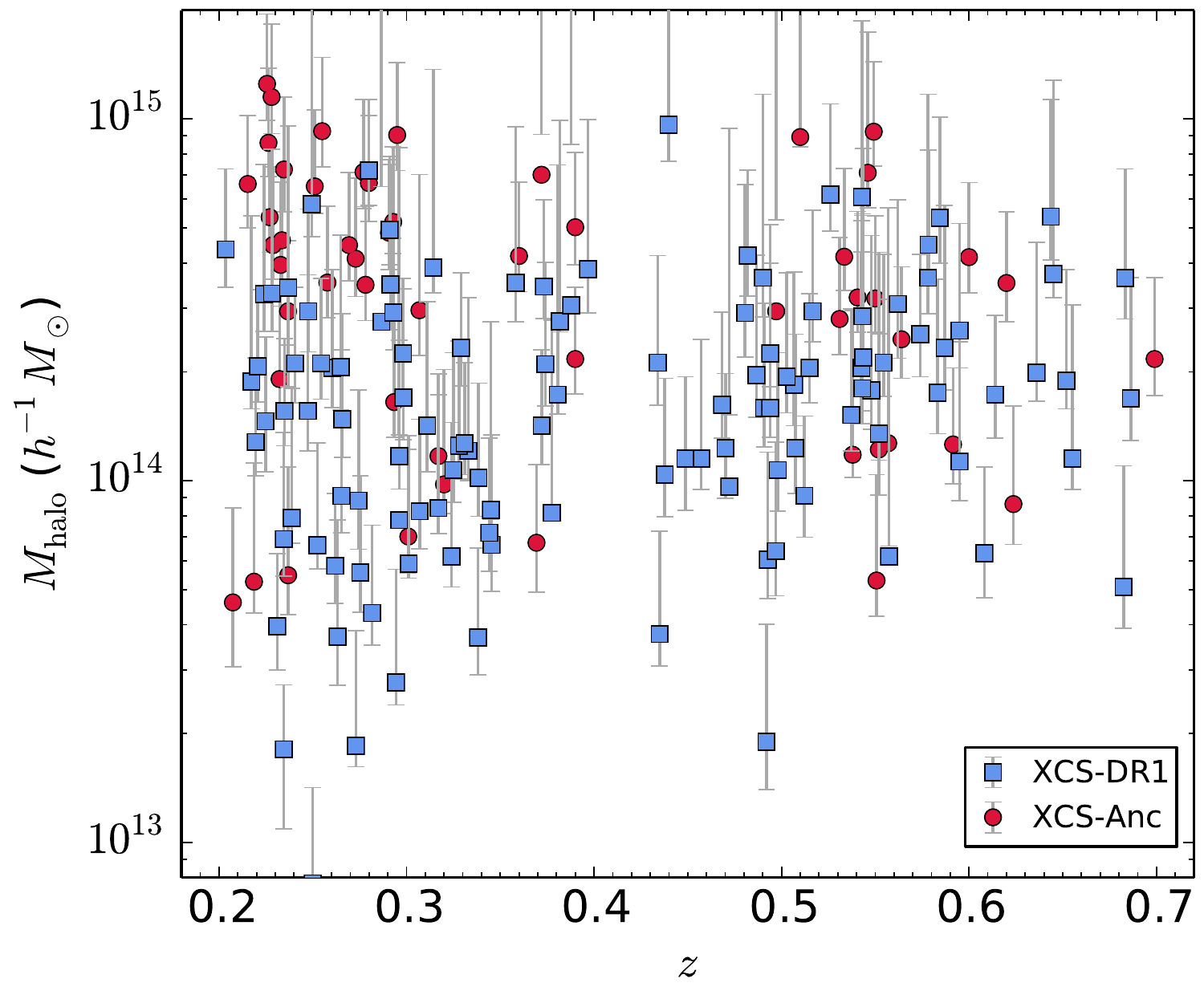}
\caption{Distribution of halo mass with spectroscopic redshift for the \TOTALCLUSTERNUMBER\, X-ray clusters used in this study. Blue symbols represent clusters in the XCS-DR1 sample, red symbols represent additional, or ``Ancillary'' clusters. The XCS-DR1 sample includes lower-mass clusters than the Ancillary sample at $z<0.5$ because high-mass clusters at $z<0.5$ are typically the intended target of their respective {\textit XMM} pointing (XCS-DR1 does not include target clusters).}
\label{Mass_z}
\end{figure}

Using these updated $z$ and $T_{\rm X}$ values, we recalculated cluster masses and radii following the method in Section~\ref{ClusterMasses}. The resulting distribution of cluster mass with redshift for our full cluster sample is shown in Fig.~\ref{Mass_z}.

\section{Measurement of the Halo Occupation Number} 
\label{Method}

We have measured the Halo Occupation Number (HON hereafter) for \TOTALCLUSTERNUMBERSpecHOD\, clusters. 
This includes \XCSinCMASSft\, clusters in the CMASS redshift range ($0.43\le z\le0.7$) and \XCSinLOWZft\, in the 
LOWZ redshift range ($0.2\le z\le0.4$). Of these, \DRSpecCMASSft\ and \ANCSpecCMASSft\ came from the XCS-DR1 
and XCS-Ancillary sample, respectively, for the CMASS sample (\DRSpecLOWZft\  and \ANCSpecLOWZft\, respectively, 
for the LOWZ sample). All of these clusters have spectroscopic redshifts, with almost all determined from BOSS 
DR11 data (only \notBOSSSpecUpdatedPercent\, per cent came from M12 or NED).

We determine the HON values by counting the number of CMASS- or LOWZ-galaxies in the vicinity of the respective cluster centroid. The method is similar to that described above (Section~\ref{betterTzs}) with regard to measuring cluster redshifts using BOSS data, i.e., we extract galaxies from a cylinder, of radius $R_{180}$, centred on the cluster location and with a depth, along the line of sight, of $\pm 3\sigma_{v}$. The HON values so derived are provided in Column 5 of Table~\ref{XCSClusterTableCMASSsmall}.

The HON values for our cluster sample are small (from \MinHODnumber\, to \MaxHODnumber), therefore it was appropriate to use Poisson uncertainties (taken from \citealt{Gehrels:1986}) to estimate the associated counting error.  
Errors on individual BOSS-galaxy redshifts will not impact the HON measurements, being $\Delta z < 0.001$, i.e., they are much smaller than the estimated $\sigma_{v}$ values. 
We have also made the simplifying assumption that the uncertainty on the mean cluster redshift is much smaller than $\sigma_{v}$. We have checked the sensitivity of the HON to the accuracy of the $R_{180}$ estimate, by recalculating the HON using both $R_{180}+1\sigma$ and $R_{180}-1\sigma$ (where the 1-$\sigma$ uncertainties in $R_{180}$ were calculated using the method described in Section~\ref{ClusterMasses}). 
For all the clusters, the HON either did not change at all, or changed less than the Poisson uncertainty, so we chose to only quote the latter in Table~\ref{XCSClusterTableCMASSsmall}, Column 5.

We present the results of our HON analysis in Table~\ref{XCSClusterTableCMASSsmall}. The column descriptions are as follows:

\begin{description}

\item[(1)] The XCS cluster ID. Encoded within each ID is the RA and Dec (J2000.0) position of the X-ray centroid. 

\item[(2)] The mean spectroscopic redshift of each cluster. The \notBOSSSpecUpdatedPercent\, per cent of clusters that came from XCS-DR1 or NED, are indicated using footnotes.

\item[(3)] An estimate of the cluster halo mass, $M_{180}$, and its 1-$\sigma$ uncertainty (see Section~\ref{ClusterMasses}). The best-fit mean halo mass is given in parentheses. We adopt the best-fit value throughout. 

\item[(4)] An estimate of the cluster virial radius, $R_{180}$, and its 1-$\sigma$ uncertainty (see Section~\ref{ClusterMasses}). The best-fit mean virial radius is given in parentheses. We adopt the best-fit value throughout.

\item[(5)] The halo occupation number, and its 1-$\sigma$ uncertainty, of BOSS-galaxies (LOWZ or CMASS).

\end{description}

\begin{table*}
\caption{The halo occupation distribution of BOSS-galaxies in XCS clusters ($0.2<z<0.7$). (A sample of 10 lines only, the full version of this table is provided via the online edition of the article). See Section~\ref{Method} for column descriptions.}
\label{XCSClusterTableCMASSsmall}
\begin{tabular}{ccccc}
\hline
\hline
XCS ID    & $z$   & $M_{180}$     & $R_{180}$    & HON  \\
(XMMXCS) & & ($10^{14} h^{-1}\,{\rm M_{\odot}}$) & ($\,h^{-1}{\rm Mpc}$) & \\
(1)    & (2) & (3)  & (4)  & (5)     \\
\hline
J022427.3$-$045028.1 & 0.490 & 7.27 $\pm$ 4.43 (3.64) & 1.91 $\pm$ 0.44 (1.60) & 6$_{-2.4}^{+3.6}$  \\
J022433.8$-$041432.9 & 0.262 & 0.69 $\pm$ 0.23 (0.58) & 0.95 $\pm$ 0.11 (0.91) & 1$_{-0.8}^{+2.3}$  \\
J022457.8$-$034851.1 & 0.614$^{1}$ & 2.09 $\pm$ 0.78 (1.73) & 1.29 $\pm$ 0.17 (1.23) & 0$_{-0.0}^{+1.8}$  \\
J022634.7$-$040408.0 & 0.345 & 1.77 $\pm$ 0.98 (0.83) & 0.84 $\pm$ 0.17 (0.68) & 1$_{-0.8}^{+2.3}$  \\
J022722.1$-$032145.2 & 0.331 & 1.58 $\pm$ 0.54 (1.27) & 1.17 $\pm$ 0.14 (1.10) & 2$_{-1.3}^{+2.6}$  \\
J022726.7$-$043209.1 & 0.307 & 1.06 $\pm$ 0.42 (0.82) & 0.93 $\pm$ 0.13 (0.87) & 1$_{-0.8}^{+2.3}$  \\
J022827.3$-$042542.2 & 0.434 & 2.91 $\pm$ 1.28 (2.12) & 1.23 $\pm$ 0.19 (1.13) & 2$_{-1.3}^{+2.6}$  \\
J023346.0$-$085048.5 & 0.265 & 1.51 $\pm$ 0.79 (0.91) & 1.09 $\pm$ 0.21 (0.95) & 1$_{-0.8}^{+2.3}$  \\
J024150.5$-$000549.9 & 0.378 & 1.79 $\pm$ 0.99 (0.82) & 1.31 $\pm$ 0.27 (1.05) & 2$_{-1.3}^{+2.6}$  \\
J025633.0$+$000558.2 & 0.360 & 5.02 $\pm$ 1.69 (4.18) & 1.84 $\pm$ 0.21 (1.75) & 4$_{-1.9}^{+3.2}$  \\
\hline
\multicolumn{5}{l} {$^1$\citet{2004A&A...423...75V}.} \\ 
\end{tabular}
\end{table*}

\section{Comparison to HOD model predictions}
\label{HODCompIntro}

Here we compare the HON of CMASS- and LOWZ-galaxies measured within XCS cluster halos (Section~\ref{Method}) to the HOD-model fits of W11 and P13. 

\begin{table*}
\caption{The mean and standard deviation of the HOD parameters as measured in \citet{White:2011} and \citet{Parejko:2013} for the CMASS and LOWZ samples, respectively. Values in parenthesis are those derived for the best-fit model (best-fit values were not reported for the LOWZ Full sample in \citealt{Parejko:2013}). 
}
\label{LitHODValues}
\begin{tabular}{ccccc}
\hline
\hline
parameter   & CMASS Full  & LOWZ NGC     & LOWZ SGC    & LOWZ Full \\
\hline
$log_{10}M_{cut}$      & 13.08 $\pm$ 0.12  (13.04)  &  13.17 $\pm$ 0.14  (13.16) & 13.09 $\pm$ 0.09  (13.11) & 13.25 $\pm$ 0.26 \\  
$log_{10}M_{1}$        & 14.06 $\pm$ 0.10  (14.05)  &  14.06 $\pm$ 0.07 (14.11)  & 14.05 $\pm$ 0.09  (14.07) & 14.18 $\pm$ 0.39 \\   
$\sigma$          & 0.98 $\pm$ 0.24    (0.94)  &   0.65  $\pm$ 0.27 (0.741)  & 0.53  $\pm$ 0.28  (0.692) & 0.70 $\pm$ 0.40 \\
$\kappa$          & 1.13 $\pm$ 0.38    (0.93)  &   1.46 $\pm$ 0.44  (0.921) & 1.74 $\pm$ 0.74   (1.26)   & 1.04 $\pm$ 0.71 \\
$\alpha$          & 0.90 $\pm$ 0.19    (0.97)  &   1.18  $\pm$ 0.18 (1.38)   & 1.31  $\pm$ 0.19  (1.31)  & 0.94 $\pm$ 0.49 \\
\hline
\end{tabular}
\end{table*}

\begin{figure*}
\begin{center}
{    
\includegraphics[scale=0.58]{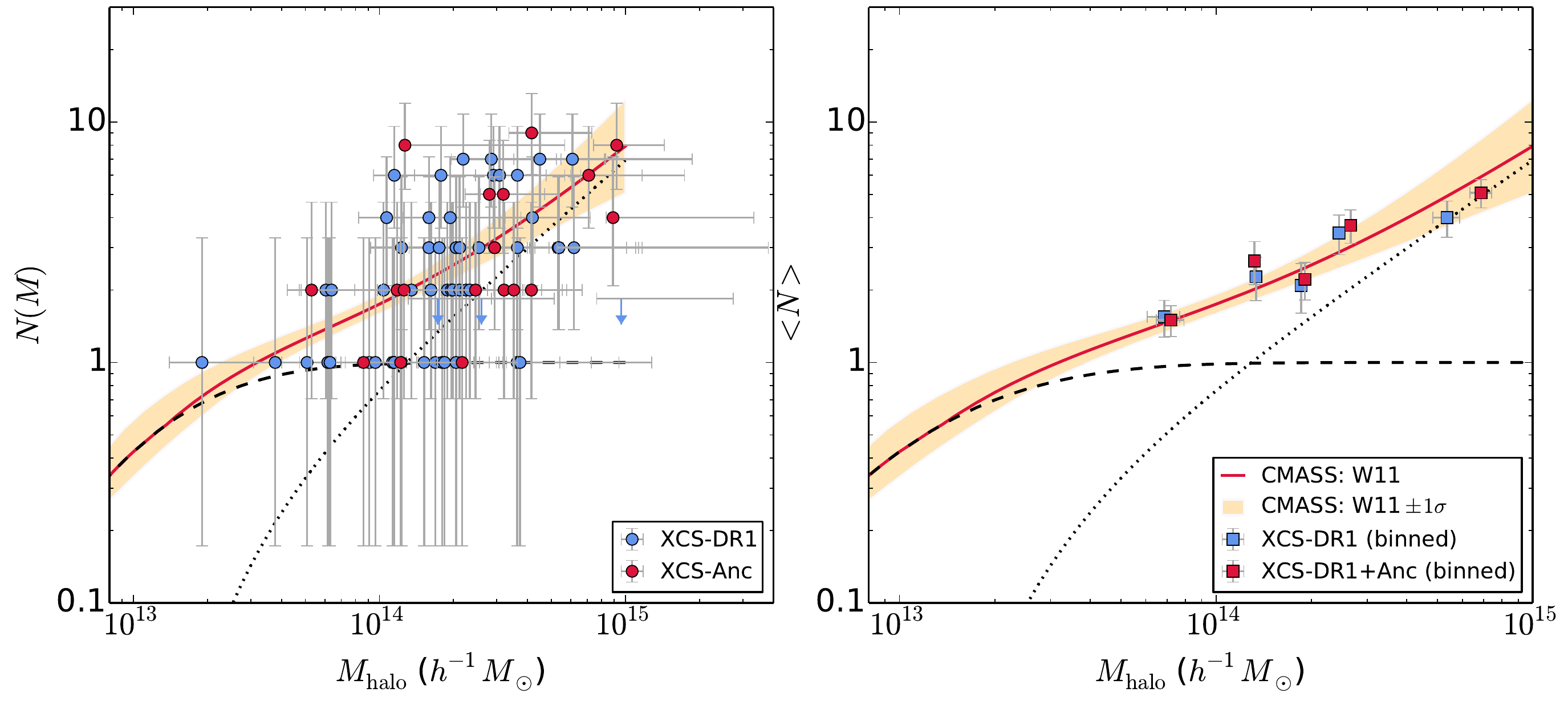}
}
\caption{LEFT: The Halo Occupation Distribution of CMASS-galaxies ($0.43 < z <0.7$) as a function of halo mass within
 \XCSinCMASSft\, X-ray selected clusters (XCS-DR1: blue circles; XCS-Ancillary: red circles). Uncertainties (including those for clusters HON value of 0) are Poisson \citep{Gehrels:1986}. For presentation purposes, points with a HON value of 0 are shown as upper limits due to the log-scale of the y-axis.
  RIGHT: The mean Halo Occupation  Distribution of CMASS-galaxies for \XCSinCMASSft\, clusters in mass bins containing approximately equal numbers of clusters (XCS-DR1: blue squares; XCS-DR1 plus XCS-Ancillary: red squares). Uncertainties on the binned points are given by the error on the mean. BOTH: The mean HOD prediction (and the 1-$\sigma$ uncertainty range) for the combined central and satellite population of W11 is indicated by the solid red line (and the yellow shaded region). The mean HOD predictions for the  separate central galaxy and satellite galaxy populations are shown by the black dashed and dotted lines, respectively. Note that the W11 results did not extend beyond $10^{15} M_{\odot}$. While the halo occupation numbers of CMASS-galaxies measured for individual clusters show a broad distribution of values, the binned values are consistent with the CMASS HOD-model fit of W11.}
\label{CMASSSpecHOD3}
\end{center}
\end{figure*}

\subsection{CMASS HOD model comparison}
\label{CMASSComp}

The left panel of Fig.~\ref{CMASSSpecHOD3} displays the HON of CMASS-galaxies measured for \XCSinCMASSft\, XCS clusters (Section~\ref{Method}). The blue symbols represent the XCS-DR1 sample only, whereas the red symbols represent the XCS-DR1 sample combined with XCS-Ancillary clusters (XCS-DR1+Anc). We also show the mean HOD-model fit to CMASS-galaxies measured from the clustering analysis of W11, along with the 1-sigma uncertainty range as given by their MCMC analysis. The data represent the HON measured for each individual cluster and therefore a broad distribution of values relative to the HOD-model (which predicts the mean HON as a function of halo mass). Nevertheless, the data show a good general agreement with the expected mean distribution. \DRSpecCMASSfthonzeroTextCaps\, XCS clusters are populated by no CMASS-galaxies (made visible by their upper limits), i.e., HON=0. These \DRSpecCMASSfthonzeroText\, X-ray selected clusters have halo masses in the range $ 0.08 - 9 \times 10^{14} h^{-1} M_{\odot}$.  We discuss the possible implications of there being massive clusters with no CMASS-galaxies in Section~\ref{fossils}. It is also noteworthy that clusters with only a single central galaxy, i.e., HON=1, are observed throughout the sampled mass range, i.e. well into the $>10^{14} h^{-1} M_{\odot}$ mass regime, see Section~\ref{fossils}.

The right panel of Fig.~\ref{CMASSSpecHOD3} presents the mean HON of CMASS-galaxies binned by cluster halo mass. The mass-range covered by each of the \MassBinsNumberText\, bins was chosen to contain (except in the case of the last bin) the same number of clusters per bin. The blue symbols represent the XCS-DR1 sample only, whereas the red symbols represent the XCS-DR1 sample combined with XCS-Ancillary clusters (XCS-DR1+Anc). There are 11 (14) XCS-DR1 (XCS-DR1+Anc) clusters per bin except in the last bin, where there are 12 (18).  Both sets of binned points (XCS-DR1 and XCS-DR1+Anc) demonstrate a clear correlation between HON and halo-mass. The uncertainty on each binned point (given by the error on the mean\footnote{Error on the mean$ = \sigma$/$\sqrt (N)$.}) overlaps with the 1-$\sigma$ uncertainty range of the HOD-model fit. This behaviour is seen for both the XCS-DR1 sample and the XCS-DR1+Anc sample. The consistency between the BOSS- and XCS-determined CMASS HOD suggests that the HOD-model fit from W11 is a reliable description of the data (Section~\ref{Discussion}).

\begin{figure*}
\begin{center}
{    
\includegraphics[scale=0.58]{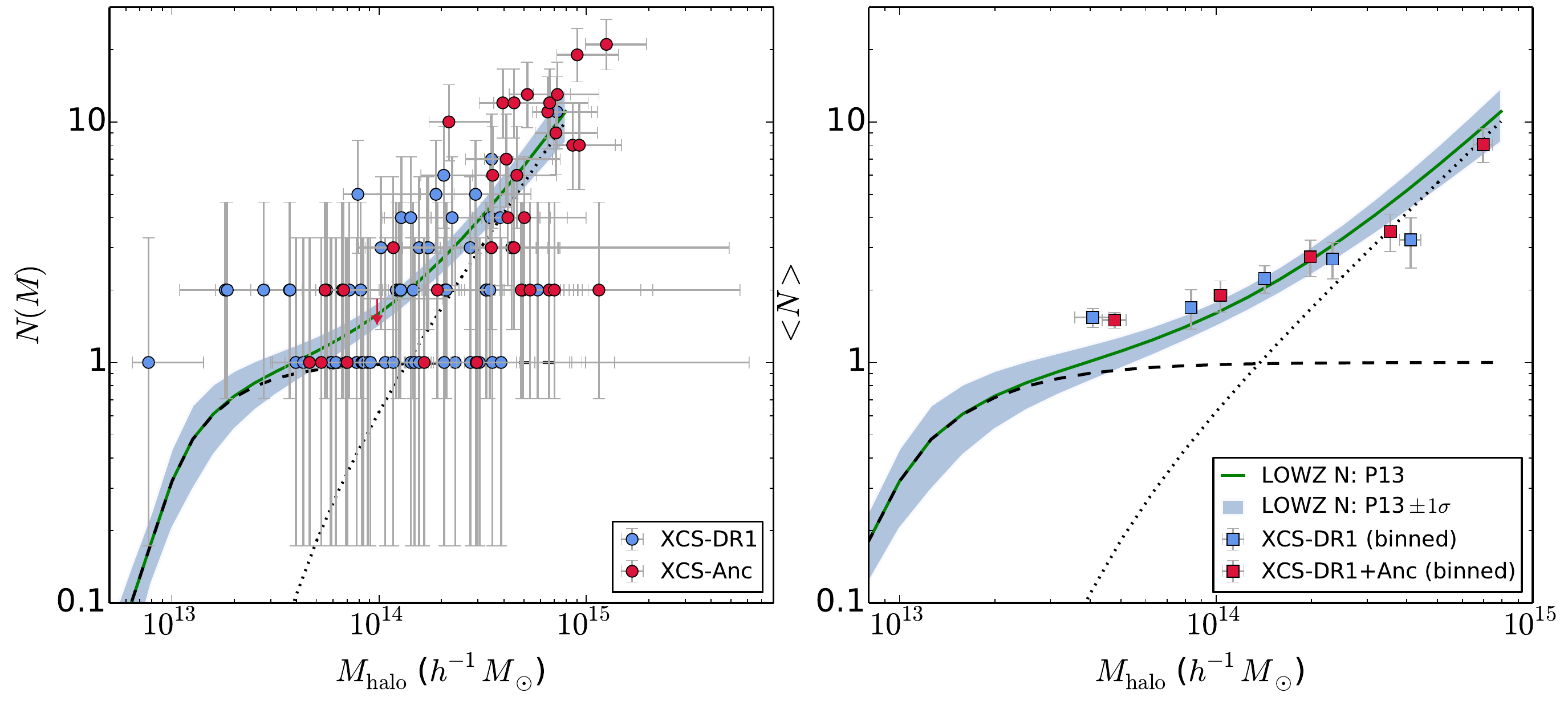}
}
\caption{LEFT: The Halo Occupation Distribution of LOWZ-galaxies ($0.2 < z <0.4$) as a function of halo mass within \XCSinLOWZft\, X-ray selected clusters (XCS-DR1: blue circles; XCS-Ancillary: red circles). Uncertainties (including those for clusters HON value of 0) are Poisson \citep{Gehrels:1986}. For presentation purposes, points with a HON value of 0 are shown as upper limits due to the log-scale of the y-axis.  RIGHT: The mean Halo Occupation Distribution of LOWZ-galaxies for \XCSinLOWZft\, clusters in mass bins chosen to contain the same number of clusters per bin. The blue squares represent the XCS-DR1 sample only, whereas the red squares represent the XCS-DR1 sample combined with XCS-Ancillary clusters (XCS-DR1+Anc). There are 13 (20) XCS-DR1 (XCS-DR1+Anc) clusters per bin, including in the last bin.  Uncertainties on the binned points are equated to the error on the mean. BOTH: The mean HOD prediction (and the 1-$\sigma$ uncertainty range) for the combined central and satellite population of P13, derived from the Northern Galactic Hemisphere, is indicated by the solid green line (and the blue shaded region). The mean HOD predictions for the  separate central galaxy and satellite galaxy populations are shown by the black dashed and dotted lines, respectively. Note that the P13 results did not extend beyond $10^{15} M_{\odot}$. While the halo occupation numbers of LOWZ-galaxies measured for individual clusters show a broad distribution of values, the binned values are consistent with the LOWZ HOD-model fit of P13.} 
\label{LOWZSpecHOD3}
\end{center}
\end{figure*}

\subsection{LOWZ HOD model comparison}
\label{LOWZComp}

The left panel of Fig.~\ref{LOWZSpecHOD3} presents the HON of LOWZ-galaxies measured within \XCSinLOWZft\, XCS clusters (Section~\ref{Method}). The right panel of Fig.~\ref{LOWZSpecHOD3} displays the mean HON of LOWZ 
galaxies binned by cluster halo mass, using the same procedure described in Section~\ref{CMASSComp}. In both panels, the blue symbols represent the XCS-DR1 sample only, whereas the red symbols represent the XCS-DR1 sample combined with XCS-Ancillary clusters (XCS-DR1+Anc). Also shown are the mean HOD-model fit to LOWZ-galaxies located over the Northern Galactic Cap (NGC) taken from P13, and the 1-$\sigma$ uncertainty range given by their MCMC analysis. Two other (to NGC) fits are presented in P13,  one to LOWZ-galaxies located over the Southern Galactic Cap (SGC), and one to a combined sample taken from both hemispheres. As shown in Table~\ref{LitHODValues}, the HOD-model fits differ between the three samples. Using a DR11 sample with higher redshift completeness than P13, \citet{Tojeiro:2014} also observe a discrepancy in the number densities and large-scale clustering power between the Northern and Southern hemispheres. They investigated a number of potential systematics that could give rise to these effects, and conclude that the excess number density observed in the SGC is most likely due to offsets in photometric calibration between the two hemispheres. In light of this tension, \citet{Tojeiro:2014} treat the NGC and SGC samples independently (and combine the clustering results from both to obtain their final BAO measurement). Given these issues, and because best-fit parameters are not provided by P13 for the combined sample, we decided to compare our HON results to the NGC model-fit only, because the NGC sample used by P13 is substantially larger than their SGC sample.

\begin{table*}
\caption{Best-fit index $\alpha$ values and 1-$\sigma$ uncertainties as inferred from the HOD of CMASS- and LOWZ-galaxies in XCS clusters.}
\label{XCSHODValues}
\begin{tabular}{lll}
\hline
\hline
Cluster sample   & CMASS $\alpha$-term   & LOWZ $\alpha$-term \\
\hline
XCS-DR1       & $0.91_{-0.11} ^{+0.10}$  ($\chi^{2}$: 67; $d.o.f.$: 54)   &    $0.98_{-0.14} ^{+0.13}$  ($\chi^{2}$: 65; $d.o.f.$: 63)  \\

XCS-DR1+Anc   & 0.91  $\pm$ 0.07  ($\chi^{2}$: 98; $d.o.f.$: 72)   &   $1.27_{-0.04} ^{+0.03}$  ($\chi^{2}$: 162; $d.o.f.$: 97) \\
\hline
\end{tabular}
\end{table*}

\subsection{A new HOD-model fit}
\label{MeanHODcomp}

The HOD-model implemented by W11 and P13 is comprised of five parameters that describe the HOD of central (Equation~\ref{eqn:ncen}) and satellite (Equation~\ref{eqn:nsat}) galaxies within dark matter halos. The sum of these two components produces the mean HOD of all galaxies within a halo of given mass (Equation~\ref{eqn:nall}). 

\begin{equation}
  N_{\rm cen}(M) = \frac{1}{2}
  \ {\rm erfc}\left[\frac{\ln(M_{\rm cut}/M)}{\sqrt{2}\sigma}\right]\,,
\label{eqn:ncen}
\end{equation}

\begin{equation}
  N_{\rm sat}(M) = N_{\rm cen}
    \left(\frac{M-\kappa M_{\rm cut}}{M_1}\right)^\alpha\,,
\label{eqn:nsat}
\end{equation}

\begin{equation}
  N(M) =  N_{\rm cen}(M)  + N_{\rm sat}(M),
\label{eqn:nall}
\end{equation}
where $M_{cut}$ is the minimum mass for a halo to host a galaxy, $M_{1}$ is the typical mass for haloes to host one satellite,  $\sigma$ is the fractional scatter in $M_{halo}$, $\kappa$ is the threshold mass for satellites and centrals to differ, and $\alpha$ is the mass dependence of the efficiency of galaxy formation.

We have explored the appropriateness of the HOD-model fits in W11 and P13 (summarised in Table~\ref{LitHODValues}) by estimating the $\alpha$-index from our measured cluster HON values. Ideally, we would have estimated all five free parameters in the HOD-model, but our data only span the mass regime pertaining to the `one-halo' term (Section~\ref{introduction}), and so are primarily sensitive to the satellite galaxy component (i.e., to the $\alpha$ parameter). Therefore, we fixed the other four parameters to best-fit values of  W11 and P13 (see Table~\ref{LitHODValues}). To determine the best-fit $\alpha$-index value, we performed a chi-squared fit, corrected for a Poisson distribution (Equation~\ref{eqn:chisq}),
\begin{equation}
  \chi^{2}_{x-p-1} = \Sigma_{x} \frac {(N_{o} - N_{e})^{2}}{N_{e}},
 \label{eqn:chisq}
\end{equation}
where $N_{o}$ is the observed halo occupation number, $N_{e}$ is the expected halo occupation number estimated from the HOD model, $x$ is the number of data points considered, and $p$ the number of degrees of freedom.

We have examined values for $\alpha$ ranging between 0.1 and 2.0 (in 0.01 steps), deemed to be a realistic representation of the data. Similar to P13, when performing our fit to the LOWZ HOD, we exclude halos that have a HON of zero. The results are shown in Table~\ref{XCSHODValues}, where the best-fit $\alpha$-index values presented correspond to the minimum chi-squared value over the $\alpha$-index range tested\footnote{Our best-fit values are close to the mean values over the parameter range tested.}. The 1-$\sigma$ uncertainty range of the $\alpha$-index value\footnote{Given by the minimum and maximum alpha-index values corresponding to one plus the minimum $\chi$-squared value, $1+\chi^2_{\rm min}$.} is also given in Table~\ref{XCSHODValues}. The $\alpha$ values presented in Table~\ref{XCSHODValues} are fully consistent with those from \citet{White:2011} and \citet{Parejko:2013} quoted in Table~\ref{LitHODValues} (see Figures~\ref{AppVFig4} and \ref{AppVFig5}). That said, the measured best-fit $\alpha$-index values vary depending on the input sample (XCS-DR1 versus XCS-DR1+Anc), and we discuss possible reasons for this result in Section~\ref{Discussion}.

\section{Discussion}
\label{Discussion}

Our aim in this paper was to examine the HOD-models for BOSS-galaxies that have been published by W11 and P13, and used in several subsequent BOSS analyses. Evidence in support of the models is provided in Figs.~\ref{CMASSSpecHOD3} and \ref{LOWZSpecHOD3}, which show our directly measured HON values to be in agreement with the model predictions, and from the slope of our CMASS HON distribution, which is consistent with the value in W11. We discuss potential sources of bias in our analysis below (Section~\ref{disc_ideas}), often drawing on the results of a comparison, photometric redshift based, CMASS HON measurement (Section~\ref{HODPhot}). We end this section with a preliminary study of HON evolution, with comparison to the predictions of \citet{Saito:2015}  (Section~\ref{HODevolv}).

\subsection{Measurement of the Halo Occupation Number using photometric redshifts}
\label{HODPhot}

We have performed an additional HOD analysis using photometric redshifts for two reasons: 1) to investigate the robustness of the results in Section~\ref{HODCompIntro}, and 2) to determine whether future HON analyses based on photometric data, for example using the Dark Energy Survey (e.g. \citealt{DES:2016}), will be reliable. For this example, we have used the proprietary photometric redshift catalogue of BOSS-targets in the CMASS redshift range ($0.4 < z < 0.7$) selected from SDSS DR8 imaging described by \citet{Ross:2011} (an equivalent photometric redshift catalogue is not available for the LOWZ BOSS-targets). The \citet{Ross:2011} analysis reproduces the CMASS target selection and measures photometric redshifts using ANNz \citep{Firth:2003} trained on 112,778 BOSS spectra acquired over the first observing semester. In addition to the CMASS colour cuts, \citet{Ross:2011} implement a seeing ($r$-band psf-FWHM $ < 2$ arcsec) and Galactic extinction (${\rm E}(B-V) < 0.08$) cut, and limit the catalogue to only cover the main SDSS DR8 imaging area. The resulting catalogue comprises 1,065,823 BOSS-targets covering an area of 9,913\,deg$^{2}$, with an estimated contamination rate (from stars and quasars) of 4.1 per cent. Star-galaxy probabilities are also assigned to each BOSS-target via ANNz, whereby a value of 1 indicates a galaxy and a value of 0 indicates a star. The RMS difference between the spectroscopic and photometric redshifts of the full training sample is $\Delta z_{photo}=0.0586$.

The SDSS DR8 footprint used by \citet{Ross:2011} covers \DRCMASSPhotft\, XCS-DR1 and \ANCCMASSPhotft\,  XCS-Ancillary clusters (all of which have spectroscopic redshifts from either M12 or NED). We have estimated the cluster parameters ($\sigma_v, T_{\rm X}, M_{180}, R_{180}, z_{\rm mean}$) for these clusters as described in Section~\ref{The data}. We have also calculated the HON of BOSS-targets for these clusters using a similar approach to that adopted in Section~\ref{Method}, although, for the association length in the transverse direction, we adopt a typical photometric redshift uncertainty ($\Delta z_{photo}=0.0586$) for each CMASS-target ($z_{cl} \pm \Delta z_{photo}$), rather than an estimate for the cluster velocity dispersion. Any given BOSS-target may turn out not to be a CMASS-galaxy, so we sum the star-galaxy probabilities of each object to generate the HON values.

Given the uncertainty on the photometric redshift estimates, corresponding to co-moving distances $\sim 10\,{\rm Mpc}$, it is necessary to perform an additional background subtraction to remove potential contamination by field galaxies. For this exercise, we estimate the typical number (as a floating point, not integer) of field galaxies falling within each cluster and subtract this estimate from the measured HON. This estimate is based on the projected $R_{180}$ area of each cluster, and the average number density of CMASS-targets in the photometric redshift catalogue in the range ($z_{cl} \pm \Delta z_{photo}$). The typical HON correction was less than 1.

To estimate 1-$\sigma$ uncertainties on the photometric HON values, we have adopted a similar, MCMC, technique to that used in Section~\ref{Method}. In this case, we only account for uncertainties in the $R_{180}$ values and hence the projected area for each cluster on the sky (as uncertainties in the transverse direction have already been considered in the statistical background subtraction). The HON are then re-calculated within the derived minimum and maximum $R_{180}$ to determine the 1-$\sigma$ range by adding a standard term for Poisson noise in quadrature.

\begin{figure*}
\begin{center}
{    
\includegraphics[scale=0.6]{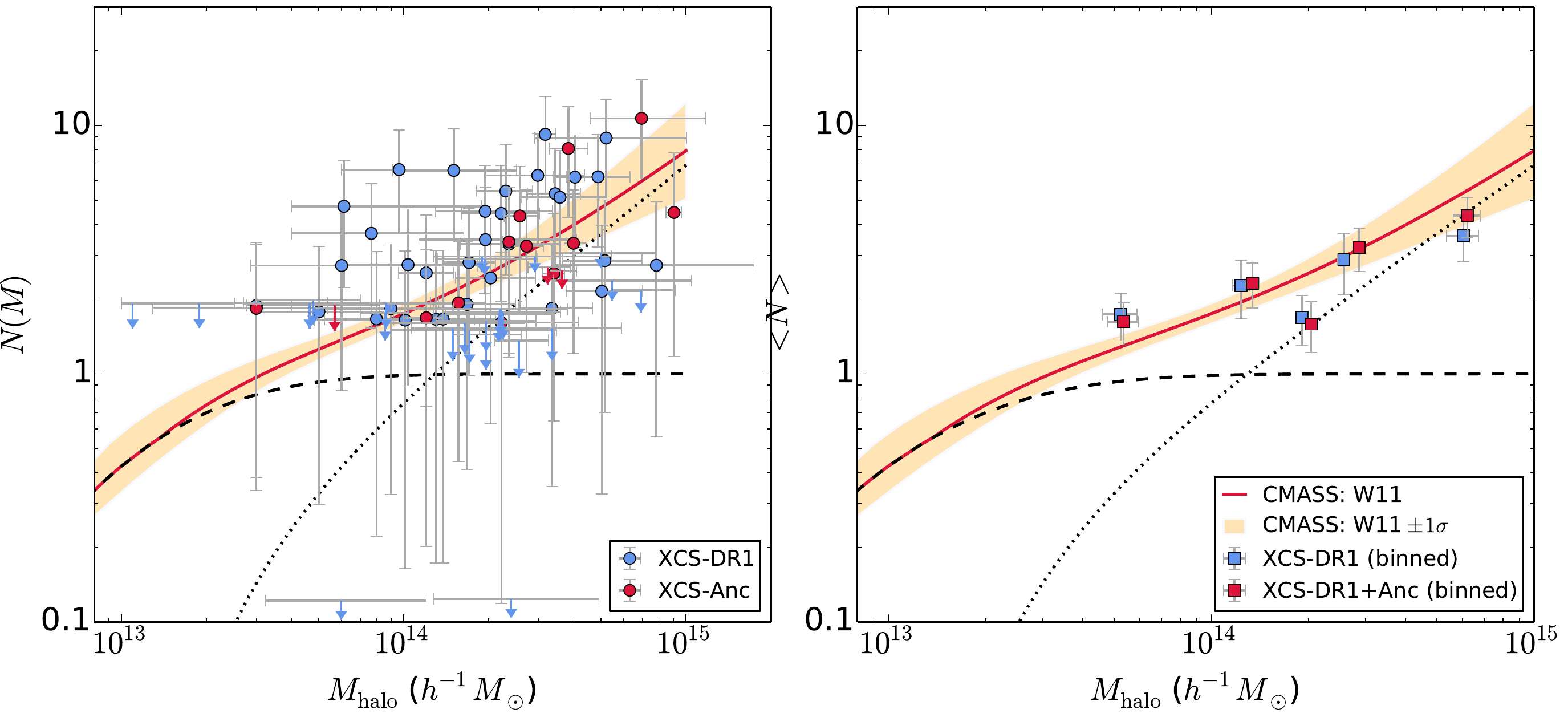}
}
\caption{LEFT: The Halo Occupation Distribution of CMASS-targets as a function of halo mass in \XCSinCMASSPhotft\, X-ray selected clusters at $0.43 < z <0.7$ (XCS-DR1: blue circles; XCS-Ancillary clusters: red circles). Points with a minimum HOD value less than 0.1 are shown as upper limits only (where the upper limit is also less than 0.1, then these are not shown at all; there are five such cases). 
RIGHT: The mean Halo Occupation Distribution of CMASS-targets for  \XCSinCMASSPhotft\, clusters in mass bins containing approximately equal numbers of clusters. Uncertainties on the binned points are set equal to the error on the mean.  BOTH: The mean HOD prediction (and the 1-$\sigma$ uncertainty range) for the combined central and satellite population of W11 is indicated by the solid red line (and the yellow shaded region). The mean HOD predictions for the  separate central galaxy and satellite galaxy populations are shown by the black dashed and dotted lines, respectively. Note that the W11 results did not extend beyond $10^{15} M_{\odot}$.
While the halo occupation numbers of CMASS-targets (measured using photometric redshift data) for individual clusters show a broad distribution of values, the binned values are consistent with the both the CMASS HOD-model fit of W11 and our measurement of the CMASS-galaxy HOD (measured using spectroscopic redshift data). This suggests our results are insensitive to BOSS redshift incompleteness. }
\label{CMASSPhotHOD1}
\end{center}
\end{figure*}

The left panel of Fig.~\ref{CMASSPhotHOD1} displays the individual HON values, which range from \MinPhotHODnumber\, to \MaxPhotHODnumber. Clusters with a HOD of zero are indicated by their 1-$\sigma$ upper limits. The right panel of Fig.~\ref{CMASSPhotHOD1} displays the mean HON of CMASS-targets binned by cluster halo mass. In both panels, the blue symbols represent the XCS-DR1 sample only, whereas the red symbols represent the XCS-DR1 sample combined with XCS-Ancillary clusters (XCS-DR1+Anc). The mass-range covered by each of the bins was chosen to contain (except in the case of the last bin) the same number of clusters per bin. There are 12 (15) XCS-DR1 (XCS-DR1+Anc) clusters per bin except in the last bin, where there are 13 (16).  Table~\ref{XCSPhotHODValues} presents the result of an HOD-model fit for CMASS-targets similar to that used to derive the Table~\ref{XCSHODValues} values for CMASS-galaxies.

\begin{table}
\caption{Best-fit index $\alpha$ values and 1-$\sigma$ uncertainties as inferred from the HOD of CMASS-targets in XCS clusters ($0.43 < z <0.7$). Fits to the XCS-DR1 and XCS-DR1 plus Ancillary clusters samples are shown, as are the minimum chi-squared value at best-fit and the number of degrees of freedom.}  
\label{XCSPhotHODValues}
\begin{tabular}{ll}
\hline
\hline
Cluster sample   & CMASS $\alpha$-term   \\
\hline
XCS-DR1       & $0.77_{-0.09} ^{+0.10}$  ($\chi^{2}$: 114; $d.o.f.$: 59)   \\
XCS-DR1+Anc   & $0.87_{-0.08} ^{+0.07}$  ($\chi^{2}$: 130; $d.o.f.$: 74)   \\
\hline
\end{tabular}
\end{table}

\subsection{Potential sources of error in our analysis}
\label{disc_ideas}

\subsubsection{Incomplete redshift information} 
\label{incompleteness}
\label{zeroHOD}
\label{fossils}

Our HON analysis has demonstrated that there are a number of genuine (i.e., confirmed by their X-ray emission) dark matter halos in the DR11 region that contain zero or one CMASS- or LOWZ-galaxies, even at masses approaching $10^{15} h^{-1} \rm M_{\odot}$. There are three cases of HON=0 in our CMASS analysis, and one in our LOWZ analysis (4 per cent and 1 per cent of the samples, respectively). There are 21 cases of HON=1 in our CMASS analysis, and 31 in our LOWZ analysis (28 per cent and 31 per cent of the samples, respectively). This  could be a reflection of the fact that the BOSS programme was incomplete in DR11, i.e., there are BOSS-targets in those HON=0,1 halos, but those had yet to be confirmed as CMASS- or LOWZ-galaxies in DR11. However, this is unlikely to be the major reason, because our investigation using the photometric redshift data (Section~\ref{HODPhot}) has shown that there are also halos without any BOSS-targets, or with only one. 

The HON=0,1 halos possibly represent `Fossil systems', i.e., systems in which the central galaxy has had time to attract and accumulate its former satellites. This hypothesis is strengthened by the fact that one of these clusters is included in the \citet{Harrison:2012} fossil system sample (the others fall outside of the \citet{Harrison:2012} redshift range, and so would not be expected to be included). It is widely accepted that fossil systems have a different evolution history, both in terms of the galaxies and the dark matter, to `normal' clusters. If genuine, these zero/low HON halos may pose a problem for the BOSS HOD-models, i.e., the model may be over predicting the mean HOD at the high mass end (if there are more massive galaxies than expected from Poisson statistics).

\subsubsection{Mismatched redshift range}

The analysis presented in W11 was conducted at an early stage of the BOSS survey. It provided an HOD-model fit to CMASS-galaxies obtained during the first-semester of BOSS observations using an early definition of the CMASS-galaxy sample. At that time, the CMASS-galaxy sample was defined to extend over the redshift range $0.4 < z < 0.7$. However, this range was later modified to $0.43 < z < 0.7$. The latter definition has been used by all subsequent BOSS analyses to constrain cosmology and investigate galaxy evolution. Therefore, in this study, we have also used $0.43 < z < 0.7$. 

\begin{figure}
\includegraphics[scale=0.55]{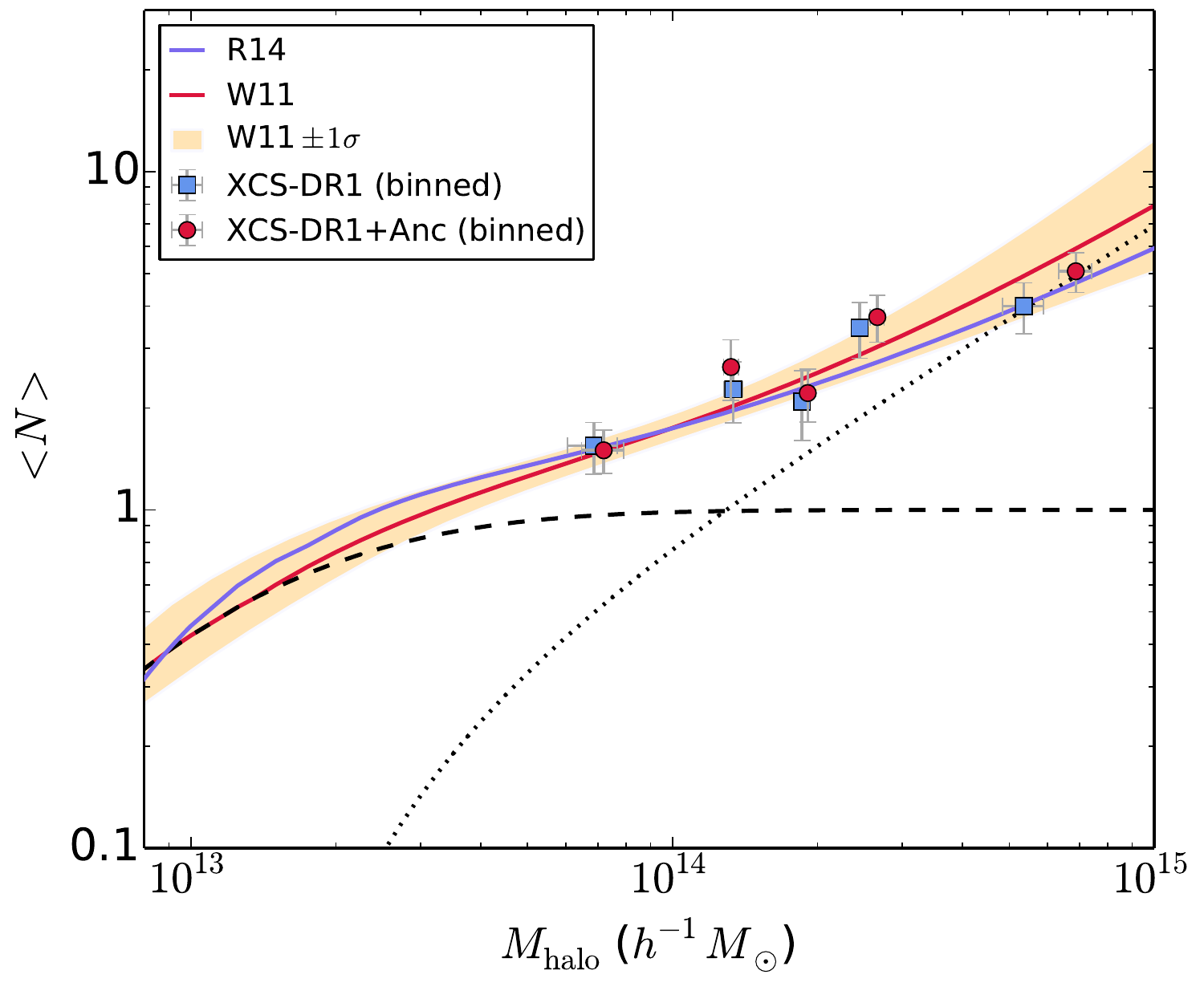}
\caption{As in Fig.~\ref{CMASSSpecHOD3} (right), but with the fiducial HOD prediction for the combined central and satellite population of \citet{Reid:2014} added (solid purple line). Both the directly-measured CMASS HOD from our study, and the CMASS HOD-model fit from W11 are consistent with this, more recent, CMASS HOD-model fit.}
\label{Comp2Reid}
\end{figure}

We have tested the impact of our adopted redshift range by repeating our CMASS analysis using the $0.4 < z < 0.7$ limits in W11. Doing so yielded an additional \NumExtraClustersOldCMASS\, clusters. Even after including those extra clusters, the best-fit $\alpha$-index values (Table~\ref{XCSHODValuespt4}) do not change significantly compared to the $0.43 < z < 0.7$ fits.  We have also re-made Fig.~\ref{CMASSSpecHOD3} (right) to include a more recent (to W11) HOD-model for CMASS-galaxies taken from \citet{Reid:2014} (Figure~\ref{Comp2Reid}) -- the \citet{Reid:2014} study uses the $0.43 < z < 0.7$ redshift range, and is consistent with both W11 and our HOD. This result suggests that if any galaxy incompleteness is present at $0.4 < z < 0.43$, it does not significantly impact the shape of the W11 HOD.

\begin{table}
\caption{Best-fit index $\alpha$ values and 1-$\sigma$ uncertainties as inferred from the HOD of CMASS-galaxies ($0.4 < z < 0.7$) in XCS clusters.}
\label{XCSHODValuespt4}
\begin{tabular}{lll}
\hline
\hline
Cluster sample   & CMASS $\alpha$-term ($0.4 < z< 0.7$)\\
\hline
XCS-DR1       & 0.84  $\pm$ 0.10  ($\chi^{2}$: 85; $d.o.f.$: 62) \\
XCS-DR1+Anc   & 0.81  $\pm$ 0.07  ($\chi^{2}$: 132; $d.o.f.$: 84)\\
\hline
\end{tabular}
\end{table}

\subsubsection{Freezing model parameters}

In our study, we have only allowed one parameter in the HOD model to vary, the slope $\alpha$. However, as shown in figure A1 of P13 (where each parameter in the model is varied separately) certain HOD parameters are degenerate to the overall shape of the correlation function. In order to include more free parameters in our fit, we would require more clusters in the HOD study, especially at the low mass end: When we tried to make a multi-parameter MCMC fit to constrain all 5 HOD parameters, the shortage of low-mass halos in our sample resulted in unconstrained fits. This, in turn, dragged the value of $\alpha$ to lower values (due to its degeneracy with $M_{1}$). Consequently, due to our inability to constrain additional HOD parameters at this stage, we report our constraints on $\alpha$ from the 1-parameter fit described in Section \ref{MeanHODcomp}.  It is hoped that a forthcoming extension of XCS will provide a sufficient number of low-mass halos to allow for more free parameters, including $M_{1}$ (i.e., the minimum halo mass required to host a satellite galaxy). We note that for the multi-parameter MCMC fit, we used the \textsc{emcee} \citep{Foreman:2013} python package, imposing uniform priors (0.01 to 2.0 on  $\alpha$, and $\pm$ 3$\sigma$ around the W11 and P13 values for the remaining HOD parameters). We also performed MCMC fits with different combinations of 3- and 4-parameters with similar outcomes.

\subsubsection{Use of cluster redshifts from the literature} 

Not all the XCS-DR1 clusters in the DR11 footprint contain one or more BOSS-galaxies within a cluster's search volume (see Section~\ref{Method}). As a consequence, it is not possible to assign these types of cluster a spectroscopic redshift using BOSS data. However, sometimes it is possible to assign a spectroscopic redshift using information in the literature.  As a result, clusters with an HON value of zero are included in our study. However, not all of the XCS clusters with an HON value of zero in the BOSS footprint are included. Those with photometric redshifts available in the literature are excluded. This is a potential source of bias, because the likelihood of a given X-ray cluster having a spectroscopic literature redshift goes up with its mass: higher mass clusters have higher X-ray fluxes (at a given redshift) and so are historically more likely to have been the target of an X-ray cluster spectroscopic follow-up campaign. Therefore, it would be worth measuring the spectroscopic redshifts of the excluded  clusters to illustrate whether our current approach has impacted the HOD-model slope.

\subsubsection{X-ray based mass determinations} 

Our analysis relies on an external normalisation for the halo mass--temperature relation based on the low redshift HIFLUGCS catalogue \citep{Pierpaoli:2001, Reiprich:2002, Viana:2003}. Not only does this approach require the extrapolation of the normalisation to higher redshifts, it also fails to take into account of the fact that measured X-ray temperature is dependent on the instrument used for the measurement (e.g. \citealt{Donahue:2014}). Independent mass measurements of the clusters, either from weak lensing or from hydrostatic mass determinations, in our study would be needed to quantify the impact of these issues.

\subsubsection{X-ray selection effects} 

It is possible that an XCS-specific selection bias is resulting in a depressed halo occupation number at the high-mass end. This is because the XCS-DR1 survey covers only a few hundred square degrees in total (albeit scattered across the BOSS footprint), meaning the volume covered at low redshifts is small compared to that of BOSS. Within this volume, many of the high mass clusters will have been the intended target of an {\em XMM} observation. As a result they will have been excluded from XCS-DR1 because target clusters are, by construct, not included in our serendipitous sample.  There is some qualitative evidence for this effect in our analysis: when ancillary clusters, which are predominately {\textit XMM} targets, are included, the averaged HONs more closely match the model predictions for the LOWZ-galaxies. In order to quantify these effects, a larger sample of XCS clusters in the BOSS footprint is needed, as is a full parameterisation of the XCS selection function.

\subsubsection{Optical selection bias} 

Another selection bias that might impact our current study arises from the optical confirmation process used in XCS-DR1. This process involved visual checks by collaboration members (at least five members per cluster) to ensure that each XCS extended source coincided with an overdensity of galaxies in optical images. The subjective nature of this process could bias the XCS-DR1 samples towards low mass clusters with higher than average HONs, hence artificially increasing the average HON in that mass range.

\subsection{Redshift evolution in the HON}
\label{HODevolv}

The recent study of \citet{Saito:2015} used subhalo abundance matching to model the stellar mass function and redshift-dependent clustering of CMASS-galaxies. Their model predicts a positive evolution in the mean-halo mass of CMASS-galaxies, which they attribute to stellar-mass incompleteness\footnote{Which they explain as fainter galaxies being missed by the magnitude cuts of the CMASS target selection.} at $z>0.6$. At higher redshifts, this effect leads to a decreasing fraction of satellite galaxies within a halo of given mass; and therefore a non-trivial variation in the HOD of CMASS-galaxies with redshift.

We investigate their prediction for a redshift-dependence of the CMASS HOD by dividing the clusters in the CMASS sample into two redshift bins at $z=0.55$. For both redshift bins, we calculate the $\alpha$-term of the CMASS HOD following the method described in Section~\ref{MeanHODcomp}. Our best-fit $\alpha$-terms are listed in Table \ref{XCSHODValuesZbins} and provide some evidence for a shallower slope on the CMASS HOD at higher redshifts, albeit with large uncertainties. This result would be expected for a decreasing fraction of satellite galaxies with redshift and lends preliminary support for the claims made in \citet{Saito:2015}.

\begin{table*}
\caption{Best-fit index $\alpha$ values and 1-$\sigma$ uncertainties as inferred from the HOD of CMASS at split into two redshift bins at $z=0.55$.}
\label{XCSHODValuesZbins}
\begin{tabular}{lll}
\hline
\hline
Cluster sample   & CMASS ($0.43 <z<0.55$) $\alpha$-term   & CMASS ($0.55 <z<0.7$) $\alpha$-term\\
\hline
XCS-DR1       & 0.96 $\pm$0.13  ($\chi^{2}$: 49; $d.o.f.$: 32)   &    $0.80_{-0.19}^{+0.18}$  ($\chi^{2}$: 17; $d.o.f.$: 20)  \\
XCS-DR1+Anc   & $0.96_{-0.09}^{+0.08}$  ($\chi^{2}$: 58; $d.o.f.$: 40)   &   $0.77_{-0.17}^{+0.16}$  ($\chi^{2}$: 40; $d.o.f.$: 30) \\
\hline
\end{tabular}
\end{table*}

\section{Conclusions}

We have performed a direct measurement of the mean HOD of BOSS-galaxies as a function of halo mass, counting the number of spectroscopically confirmed BOSS-galaxies ($0.2 < z < 0.7$) in \TOTALCLUSTERNUMBERSpecHOD\, X-ray selected galaxy clusters (${\rm log_{10}} (M_{180}/M_{\odot}) =13-15$). We have also performed a similar analysis of BOSS-targets ($0.43 < z < 0.7$) in  \XCSinCMASSPhotft\, X-ray selected galaxy clusters (there is considerable overlap between the two cluster samples at $z>0.43$). This analysis has demonstrated the following:

\begin{enumerate}

\item[(1)] When using spectroscopic redshifts from BOSS, the shape of the directly measured BOSS HOD function is consistent with the models predicted by the clustering analyses of \citet{White:2011} and \citet{Parejko:2013} for the CMASS ($0.43 < z< 0.7$) and LOWZ ($0.2 < z < 0.4$) BOSS-galaxy samples, respectively.

\item[(2)] When other parameters in the HOD model are frozen (to best-fit values of  W11 and P13), we measure best-fit slopes of $\alpha=0.91\pm 0.08$ and $\alpha=1.27^{+0.03}_{-0.04}$ (when XCS-Ancillary clusters are included) for the CMASS and LOWZ HOD, respectively. These values are consistent with the \citet{White:2011} HOD-model fit for the CMASS sample and with the \citet{Parejko:2013}  HOD-model fit for the LOWZ sample.

\item[(3)] The first two conclusions suggest the simple framework of the HOD-model is sufficient to fully describe the small-scale clustering of galaxies within halos at the galaxy-group to galaxy-cluster scale. 
  
\item[(4)] The lower $\alpha$-value of the LOWZ HOD measured from the XCS-DR-only sample (compared to the XCS-DR1 plus XCS-Anciliary sample) suggests that selection effects in the XCS-DR1 (M12) sample may be a factor, e.g. because high-mass clusters tend to be {\em XMM} targets, and hence excluded from XCS-DR1. 

\item[(5)] When using photometric redshifts that were calculated specifically for BOSS-target galaxies in the CMASS redshift range (within \TotalCMASSPhotft\, XCS clusters), we find the shape of the directly measured BOSS HOD function, and the measured slope, is consistent with the models predicted by the clustering analyses of \citet{White:2011}. 

\item[(6)] In both the spectroscopic and the photometric analysis there are examples of massive halos (where the masses are determined from their X-ray properties) that contain either one or zero BOSS-galaxies.

\item[(7)] Conclusions 5  and 6 suggest that redshift incompleteness in the SDSS-DR11 sample is not the reason why some massive (including $>10^{14} h^{-1}\,\rm{M}_{\odot}$) halos contain either one or zero BOSS-galaxies.

\item[(8)] Conclusion 5 demonstrates that it will be possible to obtain new understanding of the HOD-model using photometric galaxy surveys, such as  The Dark Energy Survey.

\item[(9)] When the redshift range of the CMASS analysis is changed from $0.43 < z< 0.7$ to $0.4 < z< 0.7$, in direct accordance with the W11 analysis, the slope ($\alpha$ value) does not change significantly. A more recent, to W11, derivation of the CMASS HOD-model \citep{Reid:2014}, was based on the $0.43 < z< 0.7$ redshift range and is similar to both our directly measured HOD and the W11 model.

\item[(10)] When the CMASS sample was divided into two redshift bins, the best-fit slope ($\alpha$ value) is shallower at $z>0.55$ compared to $z<0.55$. This result provides preliminary support to the \citet{Saito:2015} prediction that there should be a decreasing fraction of satellite galaxies within a halo of given mass.

There are several ways that our study could be improved in future. These include: including X-ray and optical selection functions to account for biases in the XCS-DR2 sample; expanding the number of free parameters in the HOD-fit; undertaking spectroscopy on a sample of XCS clusters in the BOSS footprint that were excluded from the current study because their redshifts were based on photometry only; and testing the normalisation of the mass estimation technique used for the XCS clusters by measuring masses for a sample of clusters through independent techniques, e.g. weak lensing shear or resolved X-ray spectroscopy.

\end{enumerate}

\section*{Acknowledgments}

NM acknowledges generous support from the Texas A\&M University and the George P. and Cynthia Woods Institute for Fundamental Physics and Astronomy. AKR, RN, CAC, ARL are supported by the UK Science and Technology Facilities Council (STFC) grants ST/K00090/1, ST/K00090/1, STL000652/1, ST/L005573/1, ST/M000966/1 and ST/L000644/1. MS acknowledges support by the Templeton Foundation. PR  acknowledges support from STFC and the University of Sussex Maths and Physical Sciences School. HW acknowledges support from SEPNet, the ICG and STFC. JPS acknowledges support from a Hintze Research Fellowship.  PTPV acknowledges financial support by Funda\c{c}\~{a}o para a Ci\^{e}ncia e a Tecnologia through project UID/FIS/04434/2013. We offer our thanks to members of the BOSS collaboration for their comments on the draft, including David Weinberg, Rachel Mandelbaum and Surhud More.

Funding for SDSS-III has been provided by the Alfred P. Sloan Foundation, the Participating Institutions, the National Science Foundation, and the U.S. Department of Energy Office of Science. The SDSS-III web site is http://www.sdss3.org/.

SDSS-III is managed by the Astrophysical Research Consortium for the Participating Institutions of the SDSS-III Collaboration including the University of Arizona, the Brazilian Participation Group, Brookhaven National Laboratory, Carnegie Mellon University, University of Florida, the French Participation Group, the German Participation Group, Harvard University, the Instituto de Astrofisica de Canarias, the Michigan State/Notre Dame/JINA Participation Group, Johns Hopkins University, Lawrence Berkeley National Laboratory, Max Planck Institute for Astrophysics, Max Planck Institute for Extraterrestrial Physics, New Mexico State University, New York University, Ohio State University, Pennsylvania State University, University of Portsmouth, Princeton University, the Spanish Participation Group, University of Tokyo, University of Utah, Vanderbilt University, University of Virginia, University of Washington, and Yale University.

This research has made use of the NASA/IPAC Extragalactic Database (NED) which is operated by the Jet Propulsion Laboratory, California Institute of Technology, under contract with the National Aeronautics and Space Administration. 

\bibliography{HODbib}

\begin{thebibliography}{116}
\expandafter\ifx\csname natexlab\endcsname\relax\def\natexlab#1{#1}\fi

\bibitem[{{Abazajian} {et~al.}(2005){Abazajian}, {Zheng}, {Zehavi}, {Weinberg},
  {Frieman}, {Berlind}, {Blanton}, {Bahcall}, {Brinkmann}, {Schneider} \&
  {Tegmark}}]{2005ApJ...625..613A}
{Abazajian} K., {et~al.}, 2005, \apj, 625, 613

\bibitem[{{Abbas} {et~al.}(2010){Abbas}, {de la Torre}, {Le F{\`e}vre},
  {Guzzo}, {Marinoni}, {Meneux}, {Pollo}, {Zamorani}, {Bottini}, {Garilli}, {Le
  Brun}, {Maccagni}, {Scaramella}, {Scodeggio}, {Tresse}, {Vettolani},
  {Zanichelli}, {Adami}, {Arnouts}, {Bardelli}, {Bolzonella}, {Cappi},
  {Charlot}, {Ciliegi}, {Contini}, {Foucaud}, {Franzetti}, {Gavignaud},
  {Ilbert}, {Iovino}, {Lamareille}, {McCracken}, {Marano}, {Mazure}, {Merighi},
  {Paltani}, {Pell{\`o}}, {Pozzetti}, {Radovich}, {Vergani}, {Zucca}, {Bondi},
  {Bongiorno}, {Brinchmann}, {Cucciati}, {de Ravel}, {Gregorini},
  {Perez-Montero}, {Mellier} \& {Merluzzi}}]{2010MNRAS.406.1306A}
{Abbas} U., {et~al.}, 2010, \mnras, 406, 1306

\bibitem[{{Aihara} {et~al.}(2011){Aihara}, {Allende Prieto}, {An}, {Anderson},
  {Aubourg}, {Balbinot}, {Beers}, {Berlind}, {Bickerton}, {Bizyaev}, {Blanton},
  {Bochanski}, {Bolton}, {Bovy}, {Brandt}, {Brinkmann}, {Brown}, {Brownstein},
  {Busca}, {Campbell}, {Carr}, {Chen}, {Chiappini}, {Comparat}, {Connolly},
  {Cortes}, {Croft}, {Cuesta}, {da Costa}, {Davenport}, {Dawson}, {Dhital},
  {Ealet}, {Ebelke}, {Edmondson}, {Eisenstein}, {Escoffier}, {Esposito},
  {Evans}, {Fan}, {Femen{\'{\i}}a Castell{\'a}}, {Font-Ribera}, {Frinchaboy},
  {Ge}, {Gillespie}, {Gilmore}, {Gonz{\'a}lez Hern{\'a}ndez}, {Gott}, {Gould},
  {Grebel}, {Gunn}, {Hamilton}, {Harding}, {Harris}, {Hawley}, {Hearty}, {Ho},
  {Hogg}, {Holtzman}, {Honscheid}, {Inada}, {Ivans}, {Jiang}, {Johnson},
  {Jordan}, {Jordan}, {Kazin}, {Kirkby}, {Klaene}, {Knapp}, {Kneib},
  {Kochanek}, {Koesterke}, {Kollmeier}, {Kron}, {Lampeitl}, {Lang}, {Le Goff},
  {Lee}, {Lin}, {Long}, {Loomis}, {Lucatello}, {Lundgren}, {Lupton}, {Ma},
  {MacDonald}, {Mahadevan}, {Maia}, {Makler}, {Malanushenko}, {Malanushenko},
  {Mandelbaum}, {Maraston}, {Margala}, {Masters}, {McBride}, {McGehee},
  {McGreer}, {M{\'e}nard}, {Miralda-Escud{\'e}}, {Morrison}, {Mullally},
  {Muna}, {Munn}, {Murayama}, {Myers}, {Naugle}, {Fausti Neto}, {Cuong Nguyen},
  {Nichol}, {O'Connell}, {Ogando}, {Olmstead}, {Oravetz}, {Padmanabhan},
  {Palanque-Delabrouille}, {Pan}, {Pandey}, {P{\^a}ris}, {Percival},
  {Petitjean}, {Pfaffenberger}, {Pforr}, {Phleps}, {Pichon}, {Pieri}, {Prada},
  {Price-Whelan}, {Raddick}, {Ramos}, {Reyl{\'e}}, {Rich}, {Richards}, {Rix},
  {Robin}, {Rocha-Pinto}, {Rockosi}, {Roe}, {Rollinde}, {Ross}, {Ross},
  {Rossetto}, {S{\'a}nchez}, {Sayres}, {Schlegel}, {Schlesinger}, {Schmidt},
  {Schneider}, {Sheldon}, {Shu}, {Simmerer}, {Simmons}, {Sivarani}, {Snedden},
  {Sobeck}, {Steinmetz}, {Strauss}, {Szalay}, {Tanaka}, {Thakar}, {Thomas},
  {Tinker}, {Tofflemire}, {Tojeiro}, {Tremonti}, {Vandenberg}, {Vargas
  Maga{\~n}a}, {Verde}, {Vogt}, {Wake}, {Wang}, {Weaver}, {Weinberg}, {White},
  {White}, {Yanny}, {Yasuda}, {Yeche} \& {Zehavi}}]{Aihara:2011}
{Aihara} H., {et~al.}, 2011, \apjs, 193, 29

\bibitem[{{Alam} {et~al.}(2015){Alam}, {Albareti}, {Allende Prieto}, {Anders},
  {Anderson}, {Anderton}, {Andrews}, {Armengaud}, {Aubourg}, {Bailey} \&
  et~al.}]{Alam:2015}
{Alam} S., {et~al.}, 2015, \apjs, 219, 12

\bibitem[{{Anderson} {et~al.}(2014){Anderson}, {Aubourg}, {Bailey}, {Beutler},
  {Bhardwaj}, {Blanton}, {Bolton}, {Brinkmann}, {Brownstein}, {Burden},
  {Chuang}, {Cuesta}, {Dawson}, {Eisenstein}, {Escoffier}, {Gunn}, {Guo}, {Ho},
  {Honscheid}, {Howlett}, {Kirkby}, {Lupton}, {Manera}, {Maraston}, {McBride},
  {Mena}, {Montesano}, {Nichol}, {Nuza}, {Olmstead}, {Padmanabhan},
  {Palanque-Delabrouille}, {Parejko}, {Percival}, {Petitjean}, {Prada},
  {Price-Whelan}, {Reid}, {Roe}, {Ross}, {Ross}, {Sabiu}, {Saito}, {Samushia},
  {S{\'a}nchez}, {Schlegel}, {Schneider}, {Scoccola}, {Seo}, {Skibba},
  {Strauss}, {Swanson}, {Thomas}, {Tinker}, {Tojeiro}, {Maga{\~n}a}, {Verde},
  {Wake}, {Weaver}, {Weinberg}, {White}, {Xu}, {Y{\`e}che}, {Zehavi} \&
  {Zhao}}]{2014MNRAS.441...24A}
{Anderson} L., {et~al.}, 2014, \mnras, 441, 24

\bibitem[{{Applegate} {et~al.}(2016){Applegate}, {Mantz}, {Allen}, {der
  Linden}, {Morris}, {Hilbert}, {Kelly}, {Burke}, {Ebeling}, {Rapetti} \&
  {Schmidt}}]{2016MNRAS.457.1522A}
{Applegate} D.~E., {et~al.}, 2016, \mnras, 457, 1522

\bibitem[{{Arnaud} {et~al.}(2005){Arnaud}, {Pointecouteau} \&
  {Pratt}}]{Arnaud:2005}
{Arnaud} M., {Pointecouteau} E., {Pratt} G.~W., 2005, \aap, 441, 893

\bibitem[{{Balogh} {et~al.}(2006){Balogh}, {Babul}, {Voit}, {McCarthy},
  {Jones}, {Lewis} \& {Ebeling}}]{Balogh:2006}
{Balogh} M.~L., {Babul} A., {Voit} G.~M., {McCarthy} I.~G., {Jones} L.~R.,
  {Lewis} G.~F., {Ebeling} H., 2006, \mnras, 366, 624

\bibitem[{{Berlind} \& {Weinberg}(2002)}]{BerlindWeinberg:2002}
{Berlind} A.~A., {Weinberg} D.~H., 2002, \apj, 575, 587

\bibitem[{{Berlind} {et~al.}(2003){Berlind}, {Weinberg}, {Benson}, {Baugh},
  {Cole}, {Dav{\'e}}, {Frenk}, {Jenkins}, {Katz} \&
  {Lacey}}]{2003ApJ...593....1B}
{Berlind} A.~A., {et~al.}, 2003, \apj, 593, 1

\bibitem[{{Blake} {et~al.}(2011){Blake}, {Kazin}, {Beutler}, {Davis},
  {Parkinson}, {Brough}, {Colless}, {Contreras}, {Couch}, {Croom}, {Croton},
  {Drinkwater}, {Forster}, {Gilbank}, {Gladders}, {Glazebrook}, {Jelliffe},
  {Jurek}, {Li}, {Madore}, {Martin}, {Pimbblet}, {Poole}, {Pracy}, {Sharp},
  {Wisnioski}, {Woods}, {Wyder} \& {Yee}}]{Blake:2011b}
{Blake} C., {et~al.}, 2011, \mnras, 418, 1707

\bibitem[{{Bolton} {et~al.}(2012){Bolton}, {Schlegel}, {Aubourg}, {Bailey},
  {Bhardwaj}, {Brownstein}, {Burles}, {Chen}, {Dawson}, {Eisenstein}, {Gunn},
  {Knapp}, {Loomis}, {Lupton}, {Maraston}, {Muna}, {Myers}, {Olmstead},
  {Padmanabhan}, {P{\^a}ris}, {Percival}, {Petitjean}, {Rockosi}, {Ross},
  {Schneider}, {Shu}, {Strauss}, {Thomas}, {Tremonti}, {Wake}, {Weaver} \&
  {Wood-Vasey}}]{Bolton:2012}
{Bolton} A.~S., {et~al.}, 2012, \aj, 144, 144

\bibitem[{{Borgani} {et~al.}(2004){Borgani}, {Murante}, {Springel}, {Diaferio},
  {Dolag}, {Moscardini}, {Tormen}, {Tornatore} \& {Tozzi}}]{Borgani:2004}
{Borgani} S., {et~al.}, 2004, \mnras, 348, 1078

\bibitem[{{Bryan} \& {Norman}(1998)}]{Bryan:1998}
{Bryan} G.~L., {Norman} M.~L., 1998, \apj, 495, 80

\bibitem[{{Capozzi} {et~al.}(2012{\natexlab{a}}){Capozzi}, {Collins}, {Stott}
  \& {Hilton}}]{Capozzi:2012}
{Capozzi} D., {Collins} C.~A., {Stott} J.~P., {Hilton} M., 2012{\natexlab{a}},
  \mnras, 419, 2821

\bibitem[{{Capozzi} {et~al.}(2012{\natexlab{b}}){Capozzi}, {Collins}, {Stott}
  \& {Hilton}}]{2012MNRAS.419.2821C}
{Capozzi} D., {Collins} C.~A., {Stott} J.~P., {Hilton} M., 2012{\natexlab{b}},
  \mnras, 419, 2821

\bibitem[{{Cole} {et~al.}(2005){Cole}, {Percival}, {Peacock}, {Norberg},
  {Baugh}, {Frenk}, {Baldry}, {Bland-Hawthorn}, {Bridges}, {Cannon}, {Colless},
  {Collins}, {Couch}, {Cross}, {Dalton}, {Eke}, {De Propris}, {Driver},
  {Efstathiou}, {Ellis}, {Glazebrook}, {Jackson}, {Jenkins}, {Lahav}, {Lewis},
  {Lumsden}, {Maddox}, {Madgwick}, {Peterson}, {Sutherland} \&
  {Taylor}}]{2005MNRAS.362..505C}
{Cole} S., {et~al.}, 2005, \mnras, 362, 505

\bibitem[{{Collister} \& {Lahav}(2005)}]{CollisterLahav:2005}
{Collister} A.~A., {Lahav} O., 2005, \mnras, 361, 415

\bibitem[{{Cooray}(2006)}]{Cooray:2006}
{Cooray} A., 2006, \mnras, 365, 842

\bibitem[{{Cooray} \& {Sheth}(2002)}]{CooraySheth:2002}
{Cooray} A., {Sheth} R., 2002, \physrep, 372, 1

\bibitem[{{Dark Energy Survey Collaboration} {et~al.}(2016){Dark Energy Survey
  Collaboration}, {Abbott}, {Abdalla}, {Aleksi{\'c}}, {Allam}, {Amara},
  {Bacon}, {Balbinot}, {Banerji}, {Bechtol}, {Benoit-L{\'e}vy}, {Bernstein},
  {Bertin}, {Blazek}, {Bonnett}, {Bridle}, {Brooks}, {Brunner}, {Buckley-Geer},
  {Burke}, {Caminha}, {Capozzi}, {Carlsen}, {Carnero-Rosell}, {Carollo},
  {Carrasco-Kind}, {Carretero}, {Castander}, {Clerkin}, {Collett}, {Conselice},
  {Crocce}, {Cunha}, {D'Andrea}, {da Costa}, {Davis}, {Desai}, {Diehl},
  {Dietrich}, {Dodelson}, {Doel}, {Drlica-Wagner}, {Estrada}, {Etherington},
  {Evrard}, {Fabbri}, {Finley}, {Flaugher}, {Foley}, {Fosalba}, {Frieman},
  {Garc{\'{\i}}a-Bellido}, {Gaztanaga}, {Gerdes}, {Giannantonio}, {Goldstein},
  {Gruen}, {Gruendl}, {Guarnieri}, {Gutierrez}, {Hartley}, {Honscheid}, {Jain},
  {James}, {Jeltema}, {Jouvel}, {Kessler}, {King}, {Kirk}, {Kron}, {Kuehn},
  {Kuropatkin}, {Lahav}, {Li}, {Lima}, {Lin}, {Maia}, {Makler}, {Manera},
  {Maraston}, {Marshall}, {Martini}, {McMahon}, {Melchior}, {Merson}, {Miller},
  {Miquel}, {Mohr}, {Morice-Atkinson}, {Naidoo}, {Neilsen}, {Nichol}, {Nord},
  {Ogando}, {Ostrovski}, {Palmese}, {Papadopoulos}, {Peiris}, {Peoples},
  {Percival}, {Plazas}, {Reed}, {Refregier}, {Romer}, {Roodman}, {Ross},
  {Rozo}, {Rykoff}, {Sadeh}, {Sako}, {S{\'a}nchez}, {Sanchez}, {Santiago},
  {Scarpine}, {Schubnell}, {Sevilla-Noarbe}, {Sheldon}, {Smith}, {Smith},
  {Soares-Santos}, {Sobreira}, {Soumagnac}, {Suchyta}, {Sullivan}, {Swanson},
  {Tarle}, {Thaler}, {Thomas}, {Thomas}, {Tucker}, {Vieira}, {Vikram},
  {Walker}, {Wechsler}, {Weller}, {Wester}, {Whiteway}, {Wilcox}, {Yanny},
  {Zhang} \& {Zuntz}}]{DES:2016}
{Dark Energy Survey Collaboration}, {et~al.}, 2016, \mnras, 460, 1270

\bibitem[{{Dawson} {et~al.}(2013){Dawson}, {Schlegel}, {Ahn}, {Anderson},
  {Aubourg}, {Bailey}, {Barkhouser}, {Bautista}, {Beifiori}, {Berlind},
  {Bhardwaj}, {Bizyaev}, {Blake}, {Blanton}, {Blomqvist}, {Bolton}, {Borde},
  {Bovy}, {Brandt}, {Brewington}, {Brinkmann}, {Brown}, {Brownstein}, {Bundy},
  {Busca}, {Carithers}, {Carnero}, {Carr}, {Chen}, {Comparat}, {Connolly},
  {Cope}, {Croft}, {Cuesta}, {da Costa}, {Davenport}, {Delubac}, {de Putter},
  {Dhital}, {Ealet}, {Ebelke}, {Eisenstein}, {Escoffier}, {Fan}, {Filiz Ak},
  {Finley}, {Font-Ribera}, {G{\'e}nova-Santos}, {Gunn}, {Guo}, {Haggard},
  {Hall}, {Hamilton}, {Harris}, {Harris}, {Ho}, {Hogg}, {Holder}, {Honscheid},
  {Huehnerhoff}, {Jordan}, {Jordan}, {Kauffmann}, {Kazin}, {Kirkby}, {Klaene},
  {Kneib}, {Le Goff}, {Lee}, {Long}, {Loomis}, {Lundgren}, {Lupton}, {Maia},
  {Makler}, {Malanushenko}, {Malanushenko}, {Mandelbaum}, {Manera}, {Maraston},
  {Margala}, {Masters}, {McBride}, {McDonald}, {McGreer}, {McMahon}, {Mena},
  {Miralda-Escud{\'e}}, {Montero-Dorta}, {Montesano}, {Muna}, {Myers},
  {Naugle}, {Nichol}, {Noterdaeme}, {Nuza}, {Olmstead}, {Oravetz}, {Oravetz},
  {Owen}, {Padmanabhan}, {Palanque-Delabrouille}, {Pan}, {Parejko},
  {P{\^a}ris}, {Percival}, {P{\'e}rez-Fournon}, {P{\'e}rez-R{\`a}fols},
  {Petitjean}, {Pfaffenberger}, {Pforr}, {Pieri}, {Prada}, {Price-Whelan},
  {Raddick}, {Rebolo}, {Rich}, {Richards}, {Rockosi}, {Roe}, {Ross}, {Ross},
  {Rossi}, {Rubi{\~n}o-Martin}, {Samushia}, {S{\'a}nchez}, {Sayres}, {Schmidt},
  {Schneider}, {Sc{\'o}ccola}, {Seo}, {Shelden}, {Sheldon}, {Shen}, {Shu},
  {Slosar}, {Smee}, {Snedden}, {Stauffer}, {Steele}, {Strauss}, {Streblyanska},
  {Suzuki}, {Swanson}, {Tal}, {Tanaka}, {Thomas}, {Tinker}, {Tojeiro},
  {Tremonti}, {Vargas Maga{\~n}a}, {Verde}, {Viel}, {Wake}, {Watson}, {Weaver},
  {Weinberg}, {Weiner}, {West}, {White}, {Wood-Vasey}, {Yeche}, {Zehavi},
  {Zhao} \& {Zheng}}]{Dawson:2013}
{Dawson} K.~S., {et~al.}, 2013, \aj, 145, 10

\bibitem[{{Deshpande} {et~al.}(2013){Deshpande}, {Blake}, {Bender},
  {Mahadevan}, {Terrien}, {Carlberg}, {Zasowski}, {Crepp}, {Rajpurohit},
  {Reyl{\'e}}, {Nidever}, {Schneider}, {Allende Prieto}, {Bizyaev}, {Ebelke},
  {Fleming}, {Frinchaboy}, {Ge}, {Hearty}, {Hern{\'a}ndez}, {Malanushenko},
  {Malanushenko}, {Majewski}, {Marchwinski}, {Muna}, {Oravetz}, {Pan},
  {Schiavon}, {Shetrone}, {Simmons}, {Stassun}, {Wilson} \&
  {Wisniewski}}]{APOGEE:2013}
{Deshpande} R., {et~al.}, 2013, \aj, 146, 156

\bibitem[{{Dickey} \& {Lockman}(1990)}]{Dickey:1990}
{Dickey} J.~M., {Lockman} F.~J., 1990, \araa, 28, 215

\bibitem[{{Donahue} {et~al.}(2014){Donahue}, {Voit}, {Mahdavi}, {Umetsu},
  {Ettori}, {Merten}, {Postman}, {Hoffer}, {Baldi}, {Coe}, {Czakon},
  {Bartelmann}, {Benitez}, {Bouwens}, {Bradley}, {Broadhurst}, {Ford},
  {Gastaldello}, {Grillo}, {Infante}, {Jouvel}, {Koekemoer}, {Kelson}, {Lahav},
  {Lemze}, {Medezinski}, {Melchior}, {Meneghetti}, {Molino}, {Moustakas},
  {Moustakas}, {Nonino}, {Rosati}, {Sayers}, {Seitz}, {Van der Wel}, {Zheng} \&
  {Zitrin}}]{Donahue:2014}
{Donahue} M., {et~al.}, 2014, \apj, 794, 136

\bibitem[{{Doroshkevich} {et~al.}(1978){Doroshkevich}, {Zeldovich} \&
  {Syunyaev}}]{Doroshkevich:1978}
{Doroshkevich} A.~G., {Zeldovich} Y.~B., {Syunyaev} R.~A., 1978, \sovast, 22,
  523

\bibitem[{{Dunkley} {et~al.}(2005){Dunkley}, {Bucher}, {Ferreira}, {Moodley} \&
  {Skordis}}]{Dunkley:2005}
{Dunkley} J., {Bucher} M., {Ferreira} P.~G., {Moodley} K., {Skordis} C., 2005,
  \mnras, 356, 925

\bibitem[{{Eisenstein} {et~al.}(2001){Eisenstein}, {Annis}, {Gunn}, {Szalay},
  {Connolly}, {Nichol}, {Bahcall}, {Bernardi}, {Burles}, {Castander},
  {Fukugita}, {Hogg}, {Ivezi{\'c}}, {Knapp}, {Lupton}, {Narayanan}, {Postman},
  {Reichart}, {Richmond}, {Schneider}, {Schlegel}, {Strauss}, {SubbaRao},
  {Tucker}, {Vanden Berk}, {Vogeley}, {Weinberg} \& {Yanny}}]{Eisenstein:2001}
{Eisenstein} D.~J., {et~al.}, 2001, \aj, 122, 2267

\bibitem[{{Eisenstein} {et~al.}(2011){Eisenstein}, {Weinberg}, {Agol},
  {Aihara}, {Allende Prieto}, {Anderson}, {Arns}, {Aubourg}, {Bailey},
  {Balbinot} \& et~al.}]{Eisenstein:2011}
{Eisenstein} D.~J., {et~al.}, 2011, \aj, 142, 72

\bibitem[{{Eisenstein} {et~al.}(2005){Eisenstein}, {Zehavi}, {Hogg},
  {Scoccimarro}, {Blanton}, {Nichol}, {Scranton}, {Seo}, {Tegmark}, {Zheng},
  {Anderson}, {Annis}, {Bahcall}, {Brinkmann}, {Burles}, {Castander},
  {Connolly}, {Csabai}, {Doi}, {Fukugita}, {Frieman}, {Glazebrook}, {Gunn},
  {Hendry}, {Hennessy}, {Ivezi{\'c}}, {Kent}, {Knapp}, {Lin}, {Loh}, {Lupton},
  {Margon}, {McKay}, {Meiksin}, {Munn}, {Pope}, {Richmond}, {Schlegel},
  {Schneider}, {Shimasaku}, {Stoughton}, {Strauss}, {SubbaRao}, {Szalay},
  {Szapudi}, {Tucker}, {Yanny} \& {York}}]{2005ApJ...633..560E}
{Eisenstein} D.~J., {et~al.}, 2005, \apj, 633, 560

\bibitem[{{Firth} {et~al.}(2003){Firth}, {Lahav} \& {Somerville}}]{Firth:2003}
{Firth} A.~E., {Lahav} O., {Somerville} R.~S., 2003, \mnras, 339, 1195

\bibitem[{{Foreman-Mackey} {et~al.}(2013){Foreman-Mackey}, {Hogg}, {Lang} \&
  {Goodman}}]{Foreman:2013}
{Foreman-Mackey} D., {Hogg} D.~W., {Lang} D., {Goodman} J., 2013, \pasp, 125,
  306

\bibitem[{{Fukugita} {et~al.}(1996){Fukugita}, {Ichikawa}, {Gunn}, {Doi},
  {Shimasaku} \& {Schneider}}]{Fukugita:1996}
{Fukugita} M., {Ichikawa} T., {Gunn} J.~E., {Doi} M., {Shimasaku} K.,
  {Schneider} D.~P., 1996, \aj, 111, 1748

\bibitem[{{Gehrels}(1986)}]{Gehrels:1986}
{Gehrels} N., 1986, \apj, 303, 336

\bibitem[{Gelman \& Rubin(1992)}]{Gelman:1992}
Gelman A., Rubin D.~B., 1992, Statist. Sci., 7, 457

\bibitem[{{Gunn} {et~al.}(1998){Gunn}, {Carr}, {Rockosi}, {Sekiguchi}, {Berry},
  {Elms}, {de Haas}, {Ivezi{\'c}}, {Knapp}, {Lupton}, {Pauls}, {Simcoe},
  {Hirsch}, {Sanford}, {Wang}, {York}, {Harris}, {Annis}, {Bartozek},
  {Boroski}, {Bakken}, {Haldeman}, {Kent}, {Holm}, {Holmgren}, {Petravick},
  {Prosapio}, {Rechenmacher}, {Doi}, {Fukugita}, {Shimasaku}, {Okada}, {Hull},
  {Siegmund}, {Mannery}, {Blouke}, {Heidtman}, {Schneider}, {Lucinio} \&
  {Brinkman}}]{Gunn:1998}
{Gunn} J.~E., {et~al.}, 1998, \aj, 116, 3040

\bibitem[{{Gunn} {et~al.}(2006){Gunn}, {Siegmund}, {Mannery}, {Owen}, {Hull},
  {Leger}, {Carey}, {Knapp}, {York}, {Boroski}, {Kent}, {Lupton}, {Rockosi},
  {Evans}, {Waddell}, {Anderson}, {Annis}, {Barentine}, {Bartoszek}, {Bastian},
  {Bracker}, {Brewington}, {Briegel}, {Brinkmann}, {Brown}, {Carr},
  {Czarapata}, {Drennan}, {Dombeck}, {Federwitz}, {Gillespie}, {Gonzales},
  {Hansen}, {Harvanek}, {Hayes}, {Jordan}, {Kinney}, {Klaene}, {Kleinman},
  {Kron}, {Kresinski}, {Lee}, {Limmongkol}, {Lindenmeyer}, {Long}, {Loomis},
  {McGehee}, {Mantsch}, {Neilsen}, {Neswold}, {Newman}, {Nitta}, {Peoples},
  {Pier}, {Prieto}, {Prosapio}, {Rivetta}, {Schneider}, {Snedden} \&
  {Wang}}]{Gunn:2006}
{Gunn} J.~E., {et~al.}, 2006, \aj, 131, 2332

\bibitem[{{Guo} {et~al.}(2016){Guo}, {Zheng}, {Behroozi}, {Zehavi}, {Chuang},
  {Comparat}, {Favole}, {Gottloeber}, {Klypin}, {Prada},
  {Rodr{\'{\i}}guez-Torres}, {Weinberg} \& {Yepes}}]{2016MNRAS.459.3040G}
{Guo} H., {et~al.}, 2016, \mnras, 459, 3040

\bibitem[{{Guo} {et~al.}(2015){Guo}, {Zheng}, {Zehavi}, {Behroozi}, {Chuang},
  {Comparat}, {Favole}, {Gottloeber}, {Klypin}, {Prada}, {Weinberg} \&
  {Yepes}}]{Guo:2015}
{Guo} H., {et~al.}, 2015, \mnras, 453, 4368

\bibitem[{{Halliday} {et~al.}(2004){Halliday}, {Milvang-Jensen}, {Poirier},
  {Poggianti}, {Jablonka}, {Arag{\'o}n-Salamanca}, {Saglia}, {De Lucia},
  {Pell{\'o}}, {Simard}, {Clowe}, {Rudnick}, {Dalcanton}, {White} \&
  {Zaritsky}}]{Halliday:2004}
{Halliday} C., {et~al.}, 2004, \aap, 427, 397

\bibitem[{{Harrison} {et~al.}(2012){Harrison}, {Miller}, {Richards},
  {Lloyd-Davies}, {Hoyle}, {Romer}, {Mehrtens}, {Hilton}, {Stott}, {Capozzi},
  {Collins}, {Deadman}, {Liddle}, {Sahl{\'e}n}, {Stanford} \&
  {Viana}}]{Harrison:2012}
{Harrison} C.~D., {et~al.}, 2012, \apj, 752, 12

\bibitem[{{Ho} {et~al.}(2009){Ho}, {Lin}, {Spergel} \& {Hirata}}]{Ho:2009}
{Ho} S., {Lin} Y.-T., {Spergel} D., {Hirata} C.~M., 2009, \apj, 697, 1358

\bibitem[{{Hoshino} {et~al.}(2015){Hoshino}, {Leauthaud}, {Lackner}, {Hikage},
  {Rozo}, {Rykoff}, {Mandelbaum}, {More}, {More}, {Saito} \&
  {Vulcani}}]{Hoshino:2015}
{Hoshino} H., {et~al.}, 2015, \mnras, 452, 998

\bibitem[{{Hu} \& {Kravtsov}(2003)}]{Hu:2003}
{Hu} W., {Kravtsov} A.~V., 2003, \apj, 584, 702

\bibitem[{{Kaiser}(1986)}]{Kaiser:1986}
{Kaiser} N., 1986, \mnras, 222, 323

\bibitem[{{Kollmeier} {et~al.}(2010){Kollmeier}, {Gould}, {Rockosi}, {Beers},
  {Knapp}, {Johnson}, {Morrison}, {Harding}, {Lee} \& {Weaver}}]{SEGUE:2010}
{Kollmeier} J.~A., {et~al.}, 2010, \apj, 723, 812

\bibitem[{{Kravtsov} {et~al.}(2004){Kravtsov}, {Berlind}, {Wechsler}, {Klypin},
  {Gottl{\"o}ber}, {Allgood} \& {Primack}}]{2004ApJ...609...35K}
{Kravtsov} A.~V., {Berlind} A.~A., {Wechsler} R.~H., {Klypin} A.~A.,
  {Gottl{\"o}ber} S., {Allgood} B., {Primack} J.~R., 2004, \apj, 609, 35

\bibitem[{{Kravtsov} {et~al.}(2006{\natexlab{a}}){Kravtsov}, {Vikhlinin} \&
  {Nagai}}]{2006ApJ...650..128K}
{Kravtsov} A.~V., {Vikhlinin} A., {Nagai} D., 2006{\natexlab{a}}, \apj, 650,
  128

\bibitem[{{Kravtsov} {et~al.}(2006{\natexlab{b}}){Kravtsov}, {Vikhlinin} \&
  {Nagai}}]{Kravtsov:2006}
{Kravtsov} A.~V., {Vikhlinin} A., {Nagai} D., 2006{\natexlab{b}}, \apj, 650,
  128

\bibitem[{{Leauthaud} {et~al.}(2012){Leauthaud}, {Tinker}, {Bundy}, {Behroozi},
  {Massey}, {Rhodes}, {George}, {Kneib}, {Benson}, {Wechsler}, {Busha},
  {Capak}, {Cort{\^e}s}, {Ilbert}, {Koekemoer}, {Le F{\`e}vre}, {Lilly},
  {McCracken}, {Salvato}, {Schrabback}, {Scoville}, {Smith} \&
  {Taylor}}]{2012ApJ...744..159L}
{Leauthaud} A., {et~al.}, 2012, \apj, 744, 159

\bibitem[{{Lee} {et~al.}(2011){Lee}, {Ge}, {Fleming}, {Stassun}, {Gaudi},
  {Barnes}, {Mahadevan}, {Eastman}, {Wright}, {Siverd}, {Gary}, {Ghezzi},
  {Laws}, {Wisniewski}, {Porto de Mello}, {Ogando}, {Maia}, {Nicolaci da
  Costa}, {Sivarani}, {Pepper}, {Nguyen}, {Hebb}, {De Lee}, {Wang}, {Wan},
  {Zhao}, {Chang}, {Groot}, {Varosi}, {Hearty}, {Hanna}, {van Eyken}, {Kane},
  {Agol}, {Bizyaev}, {Bochanski}, {Brewington}, {Chen}, {Costello}, {Dou},
  {Eisenstein}, {Fletcher}, {Ford}, {Guo}, {Holtzman}, {Jiang}, {French Leger},
  {Liu}, {Long}, {Malanushenko}, {Malanushenko}, {Malik}, {Oravetz}, {Pan},
  {Rohan}, {Schneider}, {Shelden}, {Snedden}, {Simmons}, {Weaver}, {Weinberg}
  \& {Xie}}]{MARVELS:2011}
{Lee} B.~L., {et~al.}, 2011, \apj, 728, 32

\bibitem[{{Lewis} \& {Bridle}(2002)}]{LewisBridle:2002}
{Lewis} A., {Bridle} S., 2002, \prd, 66, 103511

\bibitem[{{Liddle} {et~al.}(2001){Liddle}, {Viana}, {Romer} \&
  {Mann}}]{Liddle:2001}
{Liddle} A.~R., {Viana} P.~T.~P., {Romer} A.~K., {Mann} R.~G., 2001, \mnras,
  325, 875

\bibitem[{{Lin} {et~al.}(2004){Lin}, {Mohr} \& {Stanford}}]{Lin:2004}
{Lin} Y.-T., {Mohr} J.~J., {Stanford} S.~A., 2004, \apj, 610, 745

\bibitem[{{Lloyd-Davies} {et~al.}(2011){Lloyd-Davies}, {Romer}, {Mehrtens},
  {Hosmer}, {Davidson}, {Sabirli}, {Mann}, {Hilton}, {Liddle}, {Viana},
  {Campbell}, {Collins}, {Dubois}, {Freeman}, {Harrison}, {Hoyle}, {Kay},
  {Kuwertz}, {Miller}, {Nichol}, {Sahl{\'e}n}, {Stanford} \&
  {Stott}}]{Lloyd-Davies:2011}
{Lloyd-Davies} E.~J., {et~al.}, 2011, \mnras, 418, 14

\bibitem[{{Maraston} {et~al.}(2009){Maraston}, {Str{\"o}mb{\"a}ck}, {Thomas},
  {Wake} \& {Nichol}}]{Maraston:2009}
{Maraston} C., {Str{\"o}mb{\"a}ck} G., {Thomas} D., {Wake} D.~A., {Nichol}
  R.~C., 2009, \mnras, 394, L107

\bibitem[{{Masters} {et~al.}(2011){Masters}, {Maraston}, {Nichol}, {Thomas},
  {Beifiori}, {Bundy}, {Edmondson}, {Higgs}, {Leauthaud}, {Mandelbaum},
  {Pforr}, {Ross}, {Ross}, {Schneider}, {Skibba}, {Tinker}, {Tojeiro}, {Wake},
  {Brinkmann} \& {Weaver}}]{Masters:2011}
{Masters} K.~L., {et~al.}, 2011, \mnras, 418, 1055

\bibitem[{{Maughan}(2007)}]{Maughan:2007}
{Maughan} B.~J., 2007, \apj, 668, 772

\bibitem[{{Mehrtens} {et~al.}(2012){Mehrtens}, {Romer}, {Hilton},
  {Lloyd-Davies}, {Miller}, {Stanford}, {Hosmer}, {Hoyle}, {Collins}, {Liddle},
  {Viana}, {Nichol}, {Stott}, {Dubois}, {Kay}, {Sahl{\'e}n}, {Young}, {Short},
  {Christodoulou}, {Watson}, {Davidson}, {Harrison}, {Baruah}, {Smith},
  {Burke}, {Mayers}, {Deadman}, {Rooney}, {Edmondson}, {West}, {Campbell},
  {Edge}, {Mann}, {Sabirli}, {Wake}, {Benoist}, {da Costa}, {Maia} \&
  {Ogando}}]{Mehrtens:2012}
{Mehrtens} N., {et~al.}, 2012, \mnras, 423, 1024

\bibitem[{{Mewe} \& {Schrijver}(1986)}]{MeweSchrijver:1986}
{Mewe} R., {Schrijver} C.~J., 1986, \aap, 169, 178

\bibitem[{{Miyatake} {et~al.}(2015){Miyatake}, {More}, {Mandelbaum}, {Takada},
  {Spergel}, {Kneib}, {Schneider}, {Brinkmann} \&
  {Brownstein}}]{2015ApJ...806....1M}
{Miyatake} H., {et~al.}, 2015, \apj, 806, 1

\bibitem[{{More} {et~al.}(2015){More}, {Miyatake}, {Mandelbaum}, {Takada},
  {Spergel}, {Brownstein} \& {Schneider}}]{2015ApJ...806....2M}
{More} S., {Miyatake} H., {Mandelbaum} R., {Takada} M., {Spergel} D.~N.,
  {Brownstein} J.~R., {Schneider} D.~P., 2015, \apj, 806, 2

\bibitem[{{More} {et~al.}(2009){More}, {van den Bosch} \&
  {Cacciato}}]{More:2009}
{More} S., {van den Bosch} F.~C., {Cacciato} M., 2009, \mnras, 392, 917

\bibitem[{{More} {et~al.}(2011){More}, {van den Bosch}, {Cacciato}, {Skibba},
  {Mo} \& {Yang}}]{More:2011}
{More} S., {van den Bosch} F.~C., {Cacciato} M., {Skibba} R., {Mo} H.~J.,
  {Yang} X., 2011, \mnras, 410, 210

\bibitem[{{Navarro} {et~al.}(1996){Navarro}, {Frenk} \&
  {White}}]{NavarroFrenkWhite:1996}
{Navarro} J.~F., {Frenk} C.~S., {White} S.~D.~M., 1996, \apj, 462, 563

\bibitem[{{Nuza} {et~al.}(2014){Nuza}, {Kitaura}, {He{\ss}}, {Libeskind} \&
  {M{\"u}ller}}]{Nuza:2014}
{Nuza} S.~E., {Kitaura} F.-S., {He{\ss}} S., {Libeskind} N.~I., {M{\"u}ller}
  V., 2014, \mnras, 445, 988

\bibitem[{{Parejko} {et~al.}(2013){Parejko}, {Sunayama}, {Padmanabhan}, {Wake},
  {Berlind}, {Bizyaev}, {Blanton}, {Bolton}, {van den Bosch}, {Brinkmann},
  {Brownstein}, {da Costa}, {Eisenstein}, {Guo}, {Kazin}, {Maia},
  {Malanushenko}, {Maraston}, {McBride}, {Nichol}, {Oravetz}, {Pan},
  {Percival}, {Prada}, {Ross}, {Ross}, {Schlegel}, {Schneider}, {Simmons},
  {Skibba}, {Tinker}, {Tojeiro}, {Weaver}, {Wetzel}, {White}, {Weinberg},
  {Thomas}, {Zehavi} \& {Zheng}}]{Parejko:2013}
{Parejko} J.~K., {et~al.}, 2013, \mnras, 429, 98

\bibitem[{{Park} {et~al.}(2015){Park}, {Krause}, {Dodelson}, {Jain}, {Amara},
  {Becker}, {Bridle}, {Clampitt}, {Crocce}, {Fosalba}, {Gaztanaga},
  {Honscheid}, {Rozo}, {Sobreira}, {S{\'a}nchez}, {Wechsler}, {Abbott},
  {Abdalla}, {Allam}, {Benoit-L{\'e}vy}, {Bertin}, {Brooks}, {Buckley-Geer},
  {Burke}, {Carnero Rosell}, {Carrasco Kind}, {Carretero}, {Castander}, {da
  Costa}, {DePoy}, {Desai}, {Dietrich}, {Doel}, {Eifler}, {Fausti Neto},
  {Fernandez}, {Finley}, {Flaugher}, {Gerdes}, {Gruen}, {Gruendl}, {Gutierrez},
  {James}, {Kent}, {Kuehn}, {Kuropatkin}, {Lima}, {Maia}, {Marshall},
  {Melchior}, {Miller}, {Miquel}, {Nichol}, {Ogando}, {Plazas}, {Roe}, {Romer},
  {Rykoff}, {Sanchez}, {Scarpine}, {Schubnell}, {Sevilla-Noarbe},
  {Soares-Santos}, {Suchyta}, {Swanson}, {Tarle}, {Thaler}, {Vikram}, {Walker},
  {Weller} \& {Zuntz}}]{2015arXiv150705353P}
{Park} Y., {et~al.}, 2015, preprint (arXiv:1507.05353)

\bibitem[{{Parkinson} {et~al.}(2012){Parkinson}, {Riemer-S{\o}rensen}, {Blake},
  {Poole}, {Davis}, {Brough}, {Colless}, {Contreras}, {Couch}, {Croom},
  {Croton}, {Drinkwater}, {Forster}, {Gilbank}, {Gladders}, {Glazebrook},
  {Jelliffe}, {Jurek}, {Li}, {Madore}, {Martin}, {Pimbblet}, {Pracy}, {Sharp},
  {Wisnioski}, {Woods}, {Wyder} \& {Yee}}]{2012PhRvD..86j3518P}
{Parkinson} D., {et~al.}, 2012, \prd, 86, 103518

\bibitem[{{Peacock} \& {Smith}(2000)}]{Peacock:2000}
{Peacock} J.~A., {Smith} R.~E., 2000, \mnras, 318, 1144

\bibitem[{{Peebles} \& {Yu}(1970)}]{PeeblesYu:1970}
{Peebles} P.~J.~E., {Yu} J.~T., 1970, \apj, 162, 815

\bibitem[{{Percival} {et~al.}(2007){Percival}, {Cole}, {Eisenstein}, {Nichol},
  {Peacock}, {Pope} \& {Szalay}}]{2007MNRAS.381.1053P}
{Percival} W.~J., {Cole} S., {Eisenstein} D.~J., {Nichol} R.~C., {Peacock}
  J.~A., {Pope} A.~C., {Szalay} A.~S., 2007, \mnras, 381, 1053

\bibitem[{{Pierpaoli} {et~al.}(2001){Pierpaoli}, {Scott} \&
  {White}}]{Pierpaoli:2001}
{Pierpaoli} E., {Scott} D., {White} M., 2001, \mnras, 325, 77

\bibitem[{{Reid} {et~al.}(2014){Reid}, {Seo}, {Leauthaud}, {Tinker} \&
  {White}}]{Reid:2014}
{Reid} B.~A., {Seo} H.-J., {Leauthaud} A., {Tinker} J.~L., {White} M., 2014,
  \mnras, 444, 476

\bibitem[{{Reid} \& {Spergel}(2009)}]{2009ApJ...698..143R}
{Reid} B.~A., {Spergel} D.~N., 2009, \apj, 698, 143

\bibitem[{{Reiprich} \& {B{\"o}hringer}(2002)}]{Reiprich:2002}
{Reiprich} T.~H., {B{\"o}hringer} H., 2002, \apj, 567, 716

\bibitem[{{Rodr{\'{\i}}guez-Puebla} {et~al.}(2013){Rodr{\'{\i}}guez-Puebla},
  {Avila-Reese} \& {Drory}}]{2013ApJ...767...92R}
{Rodr{\'{\i}}guez-Puebla} A., {Avila-Reese} V., {Drory} N., 2013, \apj, 767, 92

\bibitem[{{Rodr{\'{\i}}guez-Torres} {et~al.}(2016){Rodr{\'{\i}}guez-Torres},
  {Chuang}, {Prada}, {Guo}, {Klypin}, {Behroozi}, {Hahn}, {Comparat}, {Yepes},
  {Montero-Dorta}, {Brownstein}, {Maraston}, {McBride}, {Tinker},
  {Gottl{\"o}ber}, {Favole}, {Shu}, {Kitaura}, {Bolton}, {Scoccimarro},
  {Samushia}, {Schlegel}, {Schneider} \& {Thomas}}]{2016MNRAS.460.1173R}
{Rodr{\'{\i}}guez-Torres} S.~A., {et~al.}, 2016, \mnras, 460, 1173

\bibitem[{{Romer} {et~al.}(2001){Romer}, {Viana}, {Liddle} \&
  {Mann}}]{Romer:2001}
{Romer} A.~K., {Viana} P.~T.~P., {Liddle} A.~R., {Mann} R.~G., 2001, \apj, 547,
  594

\bibitem[{{Ross} {et~al.}(2011){Ross}, {Ho}, {Cuesta}, {Tojeiro}, {Percival},
  {Wake}, {Masters}, {Nichol}, {Myers}, {de Simoni}, {Seo},
  {Hern{\'a}ndez-Monteagudo}, {Crittenden}, {Blanton}, {Brinkmann}, {da Costa},
  {Guo}, {Kazin}, {Maia}, {Maraston}, {Padmanabhan}, {Prada}, {Ramos},
  {Sanchez}, {Schlafly}, {Schlegel}, {Schneider}, {Skibba}, {Thomas}, {Weaver},
  {White} \& {Zehavi}}]{Ross:2011}
{Ross} A.~J., {et~al.}, 2011, \mnras, 417, 1350

\bibitem[{{Rykoff} {et~al.}(2016){Rykoff}, {Rozo}, {Hollowood},
  {Bermeo-Hernandez}, {Jeltema}, {Mayers}, {Romer}, {Rooney}, {Saro}, {Vergara
  Cervantes}, {Wechsler}, {Wilcox}, {Abbott}, {Abdalla}, {Allam}, {Annis},
  {Benoit-L{\'e}vy}, {Bernstein}, {Bertin}, {Brooks}, {Burke}, {Capozzi},
  {Carnero Rosell}, {Carrasco Kind}, {Castander}, {Childress}, {Collins},
  {Cunha}, {D'Andrea}, {da Costa}, {Davis}, {Desai}, {Diehl}, {Dietrich},
  {Doel}, {Evrard}, {Finley}, {Flaugher}, {Fosalba}, {Frieman}, {Glazebrook},
  {Goldstein}, {Gruen}, {Gruendl}, {Gutierrez}, {Hilton}, {Honscheid}, {Hoyle},
  {James}, {Kay}, {Kuehn}, {Kuropatkin}, {Lahav}, {Lewis}, {Lidman}, {Lima},
  {Maia}, {Mann}, {Marshall}, {Martini}, {Melchior}, {Miller}, {Miquel},
  {Mohr}, {Nichol}, {Nord}, {Ogando}, {Plazas}, {Reil}, {Sahl{\'e}n},
  {Sanchez}, {Santiago}, {Scarpine}, {Schubnell}, {Sevilla-Noarbe}, {Smith},
  {Soares-Santos}, {Sobreira}, {Stott}, {Suchyta}, {Swanson}, {Tarle},
  {Thomas}, {Tucker}, {Uddin}, {Viana}, {Vikram}, {Walker} \&
  {Zhang}}]{2016ApJS..224....1R}
{Rykoff} E.~S., {et~al.}, 2016, \apjs, 224, 1

\bibitem[{{Sahl{\'e}n} {et~al.}(2009){Sahl{\'e}n}, {Viana}, {Liddle}, {Romer},
  {Davidson}, {Hosmer}, {Lloyd-Davies}, {Sabirli}, {Collins}, {Freeman},
  {Hilton}, {Hoyle}, {Kay}, {Mann}, {Mehrtens}, {Miller}, {Nichol}, {Stanford}
  \& {West}}]{Sahlen:2009}
{Sahl{\'e}n} M., {et~al.}, 2009, \mnras, 397, 577

\bibitem[{{Saito} {et~al.}(2016){Saito}, {Leauthaud}, {Hearin}, {Bundy},
  {Zentner}, {Behroozi}, {Reid}, {Sinha}, {Coupon}, {Tinker}, {White} \&
  {Schneider}}]{Saito:2015}
{Saito} S., {et~al.}, 2016, \mnras, 460, 1457

\bibitem[{{S{\'a}nchez} {et~al.}(2012){S{\'a}nchez}, {Sc{\'o}ccola}, {Ross},
  {Percival}, {Manera}, {Montesano}, {Mazzalay}, {Cuesta}, {Eisenstein},
  {Kazin}, {McBride}, {Mehta}, {Montero-Dorta}, {Padmanabhan}, {Prada},
  {Rubi{\~n}o-Mart{\'{\i}}n}, {Tojeiro}, {Xu}, {Maga{\~n}a}, {Aubourg},
  {Bahcall}, {Bailey}, {Bizyaev}, {Bolton}, {Brewington}, {Brinkmann},
  {Brownstein}, {Gott}, {Hamilton}, {Ho}, {Honscheid}, {Labatie},
  {Malanushenko}, {Malanushenko}, {Maraston}, {Muna}, {Nichol}, {Oravetz},
  {Pan}, {Ross}, {Roe}, {Reid}, {Schlegel}, {Shelden}, {Schneider}, {Simmons},
  {Skibba}, {Snedden}, {Thomas}, {Tinker}, {Wake}, {Weaver}, {Weinberg},
  {White}, {Zehavi} \& {Zhao}}]{2012MNRAS.425..415S}
{S{\'a}nchez} A.~G., {et~al.}, 2012, \mnras, 425, 415

\bibitem[{{Seljak}(2000)}]{Seljak:2000}
{Seljak} U., 2000, \mnras, 318, 203

\bibitem[{{Simon} {et~al.}(2009{\natexlab{a}}){Simon}, {Hetterscheidt}, {Wolf},
  {Meisenheimer}, {Hildebrandt}, {Schneider}, {Schirmer} \&
  {Erben}}]{Simon:2009}
{Simon} P., {Hetterscheidt} M., {Wolf} C., {Meisenheimer} K., {Hildebrandt} H.,
  {Schneider} P., {Schirmer} M., {Erben} T., 2009{\natexlab{a}}, \mnras, 398,
  807

\bibitem[{{Simon} {et~al.}(2009{\natexlab{b}}){Simon}, {Hetterscheidt}, {Wolf},
  {Meisenheimer}, {Hildebrandt}, {Schneider}, {Schirmer} \&
  {Erben}}]{2009MNRAS.398..807S}
{Simon} P., {Hetterscheidt} M., {Wolf} C., {Meisenheimer} K., {Hildebrandt} H.,
  {Schneider} P., {Schirmer} M., {Erben} T., 2009{\natexlab{b}}, \mnras, 398,
  807

\bibitem[{{Skibba} {et~al.}(2014){Skibba}, {Smith}, {Coil}, {Moustakas},
  {Aird}, {Blanton}, {Bray}, {Cool}, {Eisenstein}, {Mendez}, {Wong} \&
  {Zhu}}]{2014ApJ...784..128S}
{Skibba} R.~A., {et~al.}, 2014, \apj, 784, 128

\bibitem[{{Smee} {et~al.}(2013){Smee}, {Gunn}, {Uomoto}, {Roe}, {Schlegel},
  {Rockosi}, {Carr}, {Leger}, {Dawson}, {Olmstead}, {Brinkmann}, {Owen},
  {Barkhouser}, {Honscheid}, {Harding}, {Long}, {Lupton}, {Loomis}, {Anderson},
  {Annis}, {Bernardi}, {Bhardwaj}, {Bizyaev}, {Bolton}, {Brewington}, {Briggs},
  {Burles}, {Burns}, {Castander}, {Connolly}, {Davenport}, {Ebelke}, {Epps},
  {Feldman}, {Friedman}, {Frieman}, {Heckman}, {Hull}, {Knapp}, {Lawrence},
  {Loveday}, {Mannery}, {Malanushenko}, {Malanushenko}, {Merrelli}, {Muna},
  {Newman}, {Nichol}, {Oravetz}, {Pan}, {Pope}, {Ricketts}, {Shelden},
  {Sandford}, {Siegmund}, {Simmons}, {Smith}, {Snedden}, {Schneider},
  {SubbaRao}, {Tremonti}, {Waddell} \& {York}}]{Smee:2013}
{Smee} S.~A., {et~al.}, 2013, \aj, 146, 32

\bibitem[{{Smith} {et~al.}(2016){Smith}, {Mazzotta}, {Okabe}, {Ziparo},
  {Mulroy}, {Babul}, {Finoguenov}, {McCarthy}, {Lieu}, {Bah{\'e}}, {Bourdin},
  {Evrard}, {Futamase}, {Haines}, {Jauzac}, {Marrone}, {Martino}, {May},
  {Taylor} \& {Umetsu}}]{2016MNRAS.456L..74S}
{Smith} G.~P., {et~al.}, 2016, \mnras, 456, L74

\bibitem[{{Stanek} {et~al.}(2006){Stanek}, {Evrard}, {B{\"o}hringer},
  {Schuecker} \& {Nord}}]{Stanek:2006}
{Stanek} R., {Evrard} A.~E., {B{\"o}hringer} H., {Schuecker} P., {Nord} B.,
  2006, \apj, 648, 956

\bibitem[{{Sunyaev} \& {Zeldovich}(1970)}]{SunyaevZeldovich:1970}
{Sunyaev} R.~A., {Zeldovich} Y.~B., 1970, \apss, 7, 3

\bibitem[{{Swanson} {et~al.}(2008){Swanson}, {Tegmark}, {Hamilton} \&
  {Hill}}]{Swanson:2008}
{Swanson} M.~E.~C., {Tegmark} M., {Hamilton} A.~J.~S., {Hill} J.~C., 2008,
  \mnras, 387, 1391

\bibitem[{{Tegmark} {et~al.}(2004){Tegmark}, {Strauss}, {Blanton}, {Abazajian},
  {Dodelson}, {Sandvik}, {Wang}, {Weinberg}, {Zehavi}, {Bahcall}, {Hoyle},
  {Schlegel}, {Scoccimarro}, {Vogeley}, {Berlind}, {Budavari}, {Connolly},
  {Eisenstein}, {Finkbeiner}, {Frieman}, {Gunn}, {Hui}, {Jain}, {Johnston},
  {Kent}, {Lin}, {Nakajima}, {Nichol}, {Ostriker}, {Pope}, {Scranton},
  {Seljak}, {Sheth}, {Stebbins}, {Szalay}, {Szapudi}, {Xu}, {Annis},
  {Brinkmann}, {Burles}, {Castander}, {Csabai}, {Loveday}, {Doi}, {Fukugita},
  {Gillespie}, {Hennessy}, {Hogg}, {Ivezi{\'c}}, {Knapp}, {Lamb}, {Lee},
  {Lupton}, {McKay}, {Kunszt}, {Munn}, {O'Connell}, {Peoples}, {Pier},
  {Richmond}, {Rockosi}, {Schneider}, {Stoughton}, {Tucker}, {vanden Berk},
  {Yanny} \& {York}}]{Tegmark:2004}
{Tegmark} M., {et~al.}, 2004, \prd, 69, 103501

\bibitem[{{Tinker} {et~al.}(2012{\natexlab{a}}){Tinker}, {Sheldon}, {Wechsler},
  {Becker}, {Rozo}, {Zu}, {Weinberg}, {Zehavi}, {Blanton}, {Busha} \&
  {Koester}}]{2012ApJ...745...16T}
{Tinker} J.~L., {et~al.}, 2012{\natexlab{a}}, \apj, 745, 16

\bibitem[{{Tinker} {et~al.}(2012{\natexlab{b}}){Tinker}, {Sheldon}, {Wechsler},
  {Becker}, {Rozo}, {Zu}, {Weinberg}, {Zehavi}, {Blanton}, {Busha} \&
  {Koester}}]{Tinker:2012}
{Tinker} J.~L., {et~al.}, 2012{\natexlab{b}}, \apj, 745, 16

\bibitem[{{Tinker} \& {Wetzel}(2010)}]{Tinker:2010}
{Tinker} J.~L., {Wetzel} A.~R., 2010, \apj, 719, 88

\bibitem[{{Tojeiro} {et~al.}(2012){Tojeiro}, {Percival}, {Brinkmann},
  {Brownstein}, {Eisenstein}, {Manera}, {Maraston}, {McBride}, {Muna}, {Reid},
  {Ross}, {Ross}, {Samushia}, {Padmanabhan}, {Schneider}, {Skibba},
  {S{\'a}nchez}, {Swanson}, {Thomas}, {Tinker}, {Verde}, {Wake}, {Weaver} \&
  {Zhao}}]{2012MNRAS.424.2339T}
{Tojeiro} R., {et~al.}, 2012, \mnras, 424, 2339

\bibitem[{{Tojeiro} {et~al.}(2014){Tojeiro}, {Ross}, {Burden}, {Samushia},
  {Manera}, {Percival}, {Beutler}, {Brinkmann}, {Brownstein}, {Cuesta},
  {Dawson}, {Eisenstein}, {Ho}, {Howlett}, {McBride}, {Montesano}, {Olmstead},
  {Parejko}, {Reid}, {S{\'a}nchez}, {Schlegel}, {Schneider}, {Tinker},
  {Maga{\~n}a} \& {White}}]{Tojeiro:2014}
{Tojeiro} R., {et~al.}, 2014, \mnras, 440, 2222

\bibitem[{{Valtchanov} {et~al.}(2004){Valtchanov}, {Pierre}, {Willis}, {Dos
  Santos}, {Jones}, {Andreon}, {Adami}, {Altieri}, {Bolzonella}, {Bremer},
  {Duc}, {Gosset}, {Jean} \& {Surdej}}]{2004A&A...423...75V}
{Valtchanov} I., {et~al.}, 2004, \aap, 423, 75

\bibitem[{{van den Bosch} {et~al.}(2007){van den Bosch}, {Yang}, {Mo},
  {Weinmann}, {Macci{\`o}}, {More}, {Cacciato}, {Skibba} \&
  {Kang}}]{VanDenBosch:2007}
{van den Bosch} F.~C., {et~al.}, 2007, \mnras, 376, 841

\bibitem[{{Viana} {et~al.}(2003){Viana}, {Kay}, {Liddle}, {Muanwong} \&
  {Thomas}}]{Viana:2003}
{Viana} P.~T.~P., {Kay} S.~T., {Liddle} A.~R., {Muanwong} O., {Thomas} P.~A.,
  2003, \mnras, 346, 319

\bibitem[{{Voit}(2005)}]{Voit:2005}
{Voit} G.~M., 2005, Advances in Space Research, 36, 701

\bibitem[{{Wetzel} {et~al.}(2012){Wetzel}, {Tinker} \&
  {Conroy}}]{2012MNRAS.424..232W}
{Wetzel} A.~R., {Tinker} J.~L., {Conroy} C., 2012, \mnras, 424, 232

\bibitem[{{White} {et~al.}(2011){White}, {Blanton}, {Bolton}, {Schlegel},
  {Tinker}, {Berlind}, {da Costa}, {Kazin}, {Lin}, {Maia}, {McBride},
  {Padmanabhan}, {Parejko}, {Percival}, {Prada}, {Ramos}, {Sheldon}, {de
  Simoni}, {Skibba}, {Thomas}, {Wake}, {Zehavi}, {Zheng}, {Nichol},
  {Schneider}, {Strauss}, {Weaver} \& {Weinberg}}]{White:2011}
{White} M., {et~al.}, 2011, \apj, 728, 126

\bibitem[{{Xue} \& {Wu}(2000)}]{XueWu:2000}
{Xue} Y.-J., {Wu} X.-P., 2000, \apj, 538, 65

\bibitem[{{Yang} {et~al.}(2003){Yang}, {Mo} \& {van den Bosch}}]{Yang:2003}
{Yang} X., {Mo} H.~J., {van den Bosch} F.~C., 2003, \mnras, 339, 1057

\bibitem[{{Yang} {et~al.}(2008){Yang}, {Mo} \& {van den Bosch}}]{Yang:2008}
{Yang} X., {Mo} H.~J., {van den Bosch} F.~C., 2008, \apj, 676, 248

\bibitem[{{York} {et~al.}(2000){York}, {Adelman}, {Anderson}, {Anderson},
  {Annis}, {Bahcall}, {Bakken}, {Barkhouser}, {Bastian}, {Berman}, {Boroski},
  {Bracker}, {Briegel}, {Briggs}, {Brinkmann}, {Brunner}, {Burles}, {Carey},
  {Carr}, {Castander}, {Chen}, {Colestock}, {Connolly}, {Crocker}, {Csabai},
  {Czarapata}, {Davis}, {Doi}, {Dombeck}, {Eisenstein}, {Ellman}, {Elms},
  {Evans}, {Fan}, {Federwitz}, {Fiscelli}, {Friedman}, {Frieman}, {Fukugita},
  {Gillespie}, {Gunn}, {Gurbani}, {de Haas}, {Haldeman}, {Harris}, {Hayes},
  {Heckman}, {Hennessy}, {Hindsley}, {Holm}, {Holmgren}, {Huang}, {Hull},
  {Husby}, {Ichikawa}, {Ichikawa}, {Ivezi{\'c}}, {Kent}, {Kim}, {Kinney},
  {Klaene}, {Kleinman}, {Kleinman}, {Knapp}, {Korienek}, {Kron}, {Kunszt},
  {Lamb}, {Lee}, {Leger}, {Limmongkol}, {Lindenmeyer}, {Long}, {Loomis},
  {Loveday}, {Lucinio}, {Lupton}, {MacKinnon}, {Mannery}, {Mantsch}, {Margon},
  {McGehee}, {McKay}, {Meiksin}, {Merelli}, {Monet}, {Munn}, {Narayanan},
  {Nash}, {Neilsen}, {Neswold}, {Newberg}, {Nichol}, {Nicinski}, {Nonino},
  {Okada}, {Okamura}, {Ostriker}, {Owen}, {Pauls}, {Peoples}, {Peterson},
  {Petravick}, {Pier}, {Pope}, {Pordes}, {Prosapio}, {Rechenmacher}, {Quinn},
  {Richards}, {Richmond}, {Rivetta}, {Rockosi}, {Ruthmansdorfer}, {Sandford},
  {Schlegel}, {Schneider}, {Sekiguchi}, {Sergey}, {Shimasaku}, {Siegmund},
  {Smee}, {Smith}, {Snedden}, {Stone}, {Stoughton}, {Strauss}, {Stubbs},
  {SubbaRao}, {Szalay}, {Szapudi}, {Szokoly}, {Thakar}, {Tremonti}, {Tucker},
  {Uomoto}, {Vanden Berk}, {Vogeley}, {Waddell}, {Wang}, {Watanabe},
  {Weinberg}, {Yanny}, {Yasuda} \& {SDSS Collaboration}}]{York:2000}
{York} D.~G., {et~al.}, 2000, \aj, 120, 1579

\bibitem[{{Zehavi} {et~al.}(2011){Zehavi}, {Zheng}, {Weinberg}, {Blanton},
  {Bahcall}, {Berlind}, {Brinkmann}, {Frieman}, {Gunn}, {Lupton}, {Nichol},
  {Percival}, {Schneider}, {Skibba}, {Strauss}, {Tegmark} \&
  {York}}]{Zehavi:2011}
{Zehavi} I., {et~al.}, 2011, \apj, 736, 59

\bibitem[{{Zhang} {et~al.}(2006){Zhang}, {B{\"o}hringer}, {Finoguenov},
  {Ikebe}, {Matsushita}, {Schuecker}, {Guzzo} \& {Collins}}]{Zhang:2006}
{Zhang} Y.-Y., {B{\"o}hringer} H., {Finoguenov} A., {Ikebe} Y., {Matsushita}
  K., {Schuecker} P., {Guzzo} L., {Collins} C.~A., 2006, \aap, 456, 55

\bibitem[{{Zheng} {et~al.}(2005{\natexlab{a}}){Zheng}, {Berlind}, {Weinberg},
  {Benson}, {Baugh}, {Cole}, {Dav{\'e}}, {Frenk}, {Katz} \&
  {Lacey}}]{Zheng:2005}
{Zheng} Z., {et~al.}, 2005{\natexlab{a}}, \apj, 633, 791

\bibitem[{{Zheng} {et~al.}(2005{\natexlab{b}}){Zheng}, {Berlind}, {Weinberg},
  {Benson}, {Baugh}, {Cole}, {Dav{\'e}}, {Frenk}, {Katz} \&
  {Lacey}}]{2005ApJ...633..791Z}
{Zheng} Z., {et~al.}, 2005{\natexlab{b}}, \apj, 633, 791

\bibitem[{{Zheng} {et~al.}(2007){Zheng}, {Coil} \& {Zehavi}}]{Zheng:2007}
{Zheng} Z., {Coil} A.~L., {Zehavi} I., 2007, \apj, 667, 760

\bibitem[{{Zheng} {et~al.}(2009){Zheng}, {Zehavi}, {Eisenstein}, {Weinberg} \&
  {Jing}}]{Zheng:2009}
{Zheng} Z., {Zehavi} I., {Eisenstein} D.~J., {Weinberg} D.~H., {Jing} Y.~P.,
  2009, \apj, 707, 554

\bibitem[{{Zu} \& {Mandelbaum}(2015)}]{2015MNRAS.454.1161Z}
{Zu} Y., {Mandelbaum} R., 2015, \mnras, 454, 1161

\end{thebibliography}

\appendix
\section{}

\begin{figure*}
\begin{center}
{    
\includegraphics[scale=0.58]{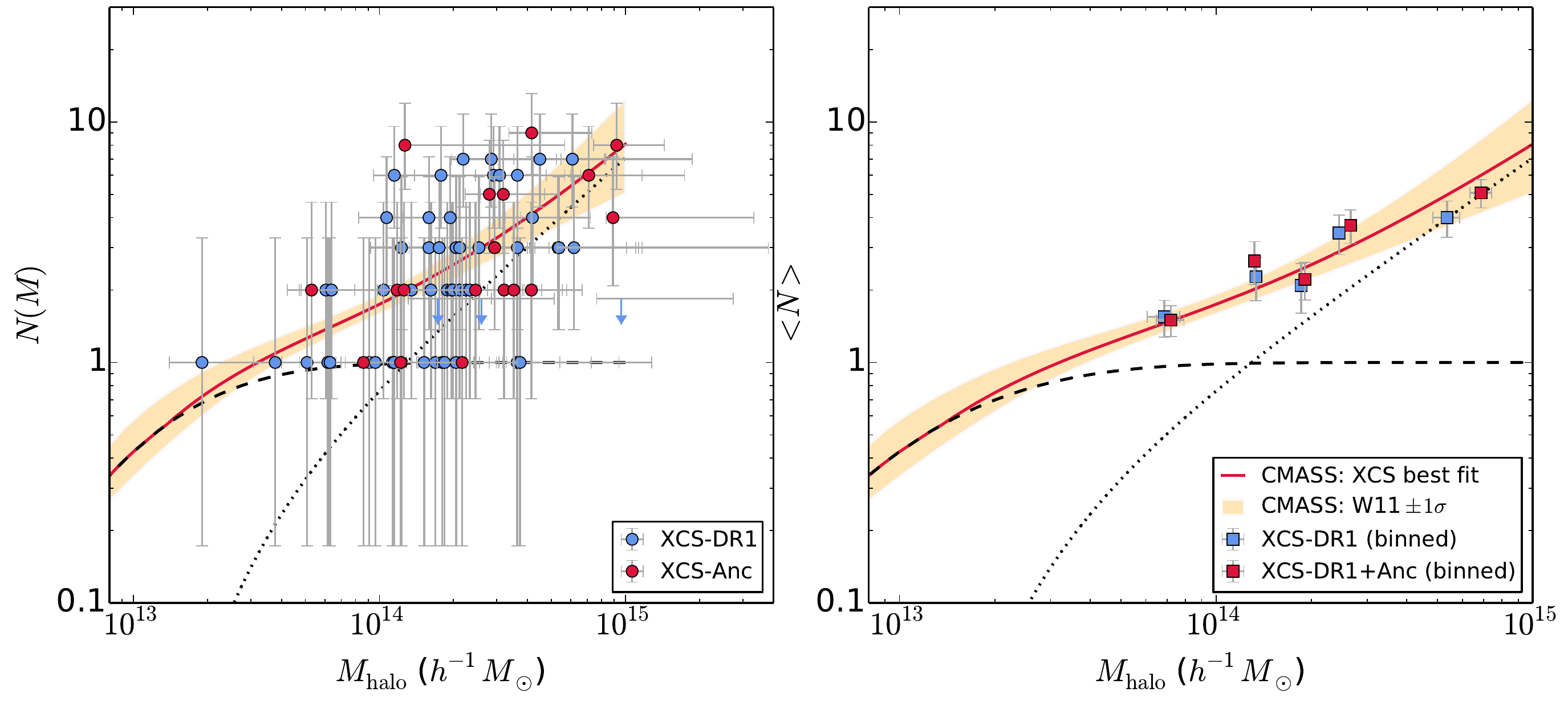}
}
\caption{As Figure \ref{CMASSSpecHOD3}, but with the XCS best fit HOD (red solid line) replacing the HOD fit from W11.}
\label{AppVFig4}
\end{center}
\end{figure*}

\begin{figure*}
\begin{center}
{    
\includegraphics[scale=0.58]{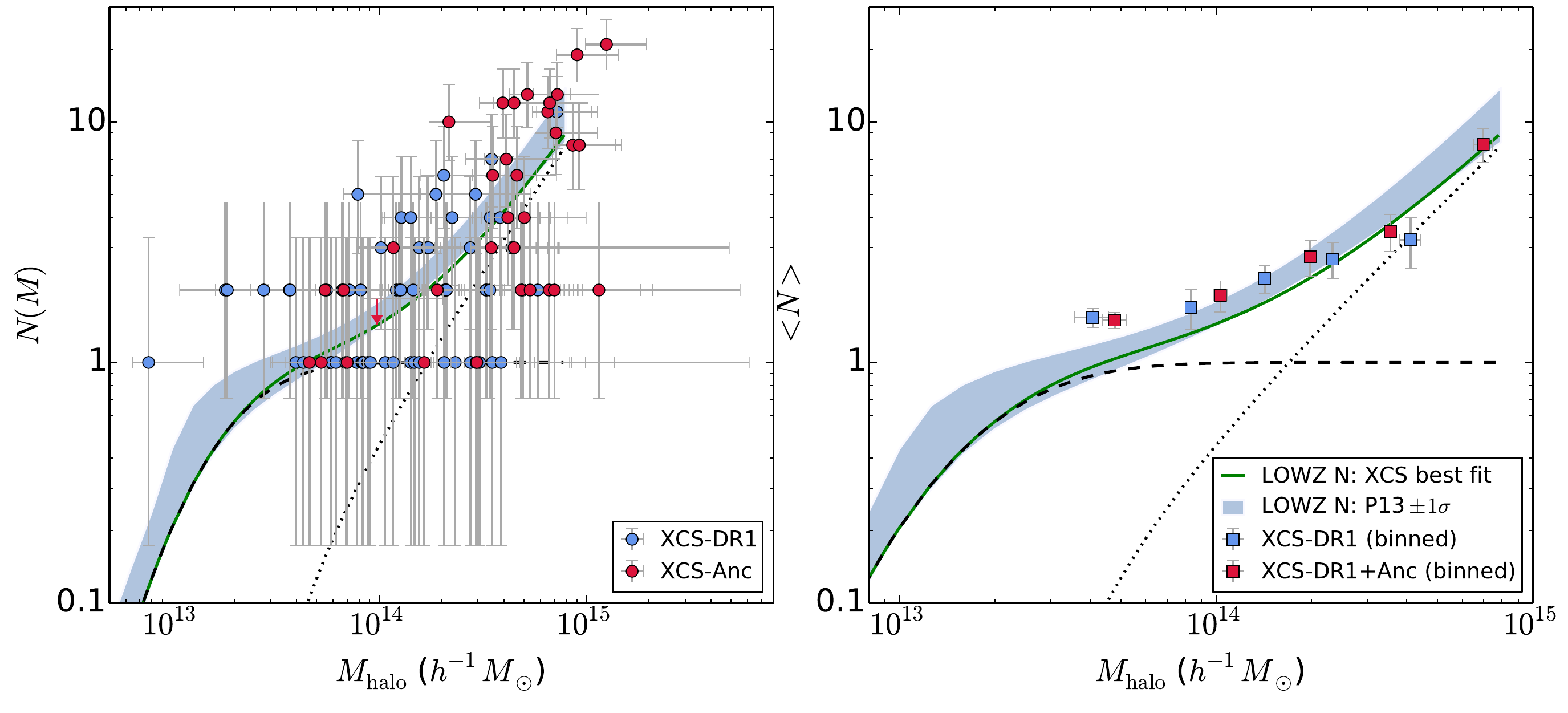}
}
\caption{As Figure \ref{LOWZSpecHOD3}, but with the XCS best fit HOD (green solid line) replacing the HOD fit from P13.} 
\label{AppVFig5}
\end{center}
\end{figure*}

\label{lastpage}

\end{document}